\documentclass[aps,pre,twocolumn]{revtex4-1}

\usepackage{amsmath}
\usepackage{amssymb}
\usepackage{color}
\usepackage{numprint}
\usepackage{physics}
\usepackage{hyperref}
\usepackage{bm}

\usepackage{graphicx}
\usepackage{changes}

\npthousandsep{\,}
\DeclareMathAlphabet{\mathitbf}{OML}{cmm}{b}{it}

\renewcommand{\th}{^{\mbox{\tiny th}}}

\newcommand{\xv}{\mathitbf x}
\newcommand{\vv}{\mathitbf v}

\newcommand{\dbar}{{\,\mathchar'26\mkern-12mu d}}

\newcommand{\sFrac}[2]{{\textstyle\frac{#1}{#2}}}

\begin{document}
\title{Fast generation of ultrastable computer glasses by minimization of an augmented potential energy}
\author{Geert Kapteijns${}^{1}$, Wencheng Ji${}^{2}$, Carolina Brito${}^{2,3}$, Matthieu Wyart${}^{2}$ and Edan Lerner${}^{1}$}
\affiliation{${}^1$ Institute for Theoretical Physics, University of Amsterdam, Science Park 904, 1098 XH Amsterdam, The Netherlands \\ ${}^2$Institute of Physics, EPFL, CH-1015 Lausanne, Switzerland \\ ${}^3$ Instituto de F\'{i}sica, UFRGS, 91501-970, Porto Alegre, Brazil}

\begin{abstract}
We present a model and protocol that enable the generation of extremely stable computer glasses at minimal computational cost. The protocol consists of an instantaneous quench in an augmented potential energy landscape, with particle radii as additional degrees of freedom. We demonstrate how our glasses' mechanical stability, which is readily tunable in our approach, is reflected both in microscopic and macroscopic observables. Our observations indicate that the stability of our computer glasses is at least comparable to that of computer glasses generated by the celebrated Swap Monte Carlo algorithm. Strikingly, some key properties support even qualitatively enhanced stability in our scheme: the density of quasilocalized excitations displays a gap in our most stable computer glasses, whose magnitude scales with the polydispersity of the particles.  We explain this observation, which is consistent with the lack of plasticity we observe at small stress. It also suggests that these glasses are depleted from two-level systems, similarly to experimental vapor-deposited ultrastable glasses.  
\end{abstract}

\maketitle

\section{introduction}
\label{introduction}

One of the key challenges in glass physics is understanding the large variations of the thermodynamic, micro- and macro-mechanical properties that glassy solids often display, depending on the protocol by which they are formed. Pronounced examples of this dependence are seen in metallic glasses: their toughness can depend in a complex manner on the degree of annealing of the pre-deformed samples \cite{fracture_toughness_eran_2018,Garrett10257}, a phenomenon attributed to `annealing embrittlement' \cite{Garrett10257}. In numerical simulations of nanoindentation of a model metallic glass it was observed that the propensity for strain localization in the form of shear banding is substantially enhanced by deeper annealing of the pre-deformed glassy samples \cite{SHI20074317}. In computer glasses made by quenching equilibrium supercooled configurations of various temperatures, it was observed that the frequencies of soft quasilocalized modes increase significantly for more deeply supercooled parent equilibrium states \cite{cge_paper}, while the spatial extent of those modes decreases \cite{inst_note}. The low-temperature thermodynamics of vapor-deposited ultrastable glasses provide another striking example of the effects of preparation protocol: the temperature dependence of their specific heat resembles that of crystalline solids \cite{ultrastable_perez_pnas,ediger_2017} instead of the ubiquitous anomalous dependence that is generically observed in glassy solids \cite{pohl_1971,Anderson}.

A recent groundbreaking advancement in computer simulations of supercooled liquids has made it possible to equilibrate supercooled liquids down to extremely low temperatures, surpassing even experimentally accessible supercooling temperature ranges \cite{berthier_prx}. This breakthrough has been achieved by employing the Swap Monte Carlo algorithm \cite{TSAI1978465, GAZZILLO1989388, Grigera2001}, and --- building on previous observations made in \cite{smarajit_epl_2015} for a three-component mixture --- carefully tayloring a model glass former such that the efficiency of the algorithm is maximized, while ensuring that the model remains robust against crystallization or fractionation.
Computer glasses formed via this computational approach display very large variations in their transient elasto-plastic response. In particular, a phase transition manifested by the nucleation of a system-spanning shear band in deformed samples is observed, depending on the temperature from which the initial, undeformed glassy samples were quenched \cite{yielding_LB_2018}. Furthermore, a study of the vibrational spectra of Swap Monte Carlo computer glasses revealed that the density of quasilocalized vibrational modes, previously shown to follow a universal non-Debye distribution $D(\omega)\!\sim\!\omega_g^{-5}\omega^4$ \cite{modes_prl,geert_non_debye}, is reduced with deep supercooling --- it retains the same power-law behavior, but the coefficient $\omega_g^{-5}$ diminishes severalfold \cite{modes_ultra_stable_LB}.

While the Swap Monte Carlo approach allows one to generate computer glasses with unprecedented stability, the accessible system sizes are inevitably limited by slow glassy dynamics at very deep supercooling. In this work we describe a computational approach, proposed by some of us in \cite{swap_jamming_prx_2018}, that consists of a direct minimization of an augmented potential energy in which the particle radii are included as additional degrees of freedom. This approach enables the generation of computer glasses that are as stable as those created via Swap Monte Carlo, at a small fraction of the computational cost.  We study the structure, micro- and macro-mechanical properties of our  computer glasses, demonstrating the large variation in glass stability that our approach provides. On the practical side, the computational speed-up offered by our approach will enable extensive statistical analyses of large ensembles of glassy samples. Using our approach very large and stable glassy samples can be generated, which will likely be useful for computational studies of transient dynamics and shear-banding instabilities under external deformations. On the physical side, our key finding is that the ultra-stable glasses we generate have a gap in their density of quasilocalized excitations: the behavior $D(\omega)\!\sim\!\omega_g^{-5}\omega^4$ breaks down below some frequency scale $\omega_{\mbox{\tiny$\Delta$}}$. We show that $\omega_{\mbox{\tiny$\Delta$}}\!\sim\!\Delta^{1/2}$, where $\Delta$ characterizes the amount of polydispersity. This result rationalizes why there is so little pre-yielding plasticity in these glasses, and suggests that they are also deprived of two-level-systems, in consistence with recent empirical observations in vapor-deposited ultrastable glasses \cite{ultrastable_perez_pnas,ediger_2017}.

This paper is organized as follows; we first provide a detailed description of the computer model employed and the protocol by which we created glassy samples in Sect.~\ref{model_and_methods}. In Sect.~\ref{structure_and_elasticity}, we present various micro- and macro-structural analyses of our computer glasses, including an analysis of the vibrational spectra. We present results from athermal quasistatic shear deformation of our computer glasses in Sect.~\ref{elasto_plastic_transients}. We rationalize the scaling of the frequency gap featured by quasilocalized excitations in Sect.\ref{discussion}.

\section{Model description and numerical procedures}
\label{model_and_methods}

In this Section we describe the computer glass model employed, and the procedure used to generate  ultra-stable glassy samples. Details about the observables measured and presented in our work can be found in Appendix~\ref{observables_appendix}, while in Appendix~\ref{aqs} we explain the athermal, quasistatic deformation protocol that we used.

\subsection{Model}

A polydisperse liquid of particles in three dimensions is usually thought as having three degrees of freedom per particle, and the particle radii or effective sizes are considered to be frozen parameters. However, in an equivalent description all particles are identical (or, in our work, come in two species to suppress nucleation) but their radii are degrees of freedom subjected to a  chemical potential, chosen so as to reproduce the desired polydispersity \cite{Glandt_1984,Glandt_1987,swap_jamming_prx_2018}. Inspired by this description, we consider a system of $N$ particles in three dimensions (3D) that interact via the potential energy
\begin{equation}
\label{111}
{\cal U} = \sum_{i<j}\varphi(r_{ij},\lambda_i,\lambda_j) + \sum_i\mu(\lambda_{i},\lambda_{i}^{(0)})\,,
\end{equation}
where $r_{ij}$ is the distance between the $i\th$ and $j\th$ particles, and $\lambda_i$ is the $i\th$ particle's effective size.  During the preparation of our glasses, the particles' effective sizes are considered to be degrees of freedom, as explained in what follows. For the pairwise interactions we use a modified Lennard-Jones (LJ) potential, which reads
\begin{widetext}
\begin{equation}\label{pairwise_potential}
\varphi(r_{ij},\lambda_i,\lambda_j) = \left\{ \begin{matrix} 6\,\varepsilon\left(\big(\sFrac{\lambda_{ij}}{r_{ij}}\big)^{12}-\big(\sFrac{\lambda_{ij}}{r_{ij}}\big)^{6} + \sum\limits_{\ell=0}^{3} c_{2\ell}\big(\sFrac{r_{ij}}{\lambda_{ij}}\big)^{2\ell}\right)\!\!\!\!\vspace{0.2cm} &, & \sFrac{r_{ij}}{\lambda_{ij}}\! <\! x_c \\ 0\!\!\!\! &,& \sFrac{r_{ij}}{\lambda_{ij}}\! \ge\! x_c \end{matrix} \right.,
\end{equation}
\end{widetext}
where $\varepsilon$ is a microscopic energy scale, $\lambda_{ij}\!\equiv\! \lambda_i\!+\!\lambda_j$, and the coefficients $c_{2\ell}$ are determined by requiring  that three derivatives of $\varphi$ with respect to the interparticle distance vanish continuously at the dimensionless cutoff $x_c$. For the sake of computational efficiency we chose $x_c\!=\!2.0$, instead of the traditional $x_c^{\mbox{\tiny LJ}}\!=\!2.5$ \cite{kablj}. 

\begin{figure}[!ht]
\centering
\includegraphics[width = 0.50\textwidth]{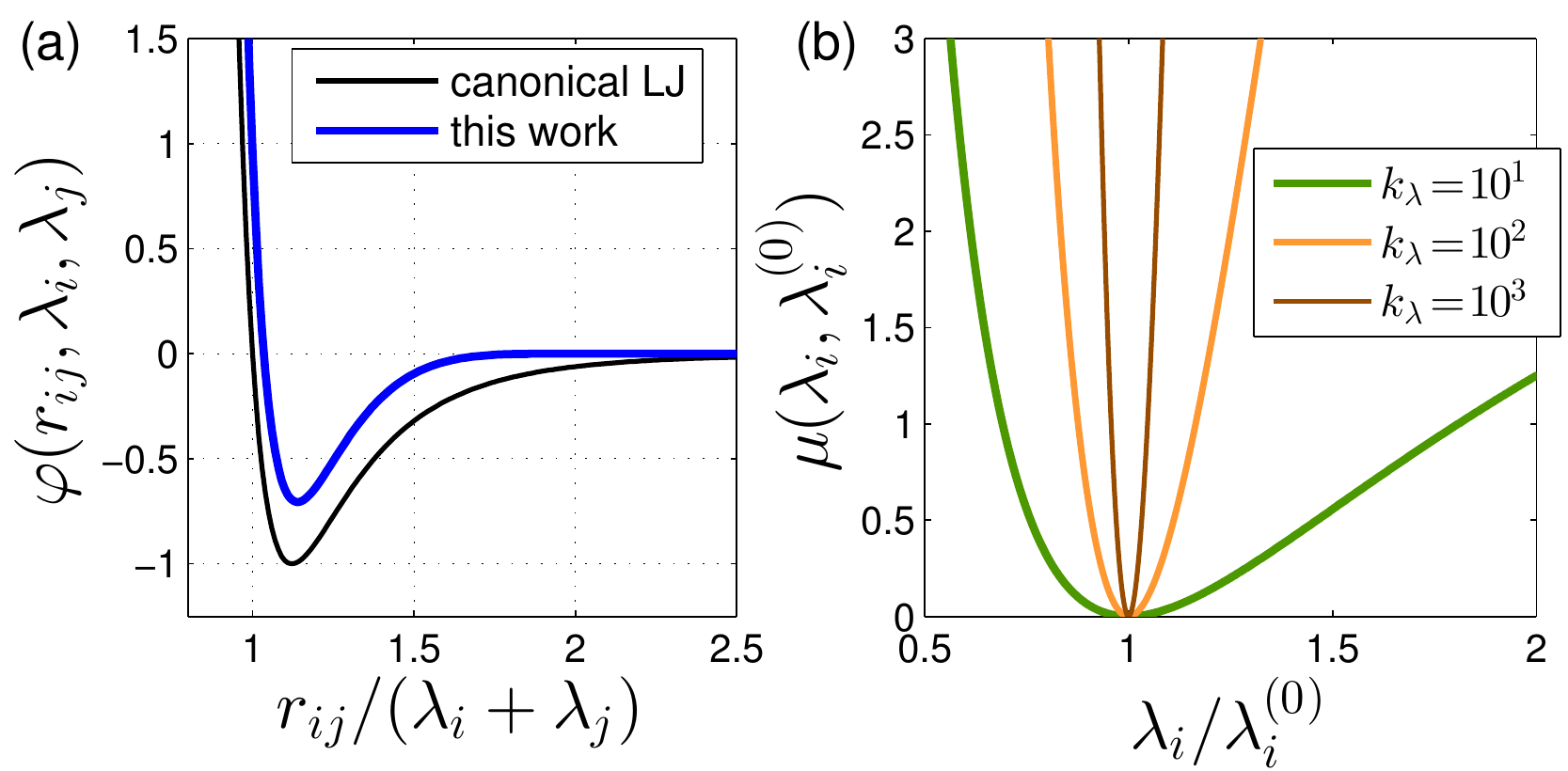}
\caption{\footnotesize (a) Pairs of nearby particles in our model glass former interact via the pairwise potential $\varphi(r_{ij},\lambda_i,\lambda_j)$ as given by Eq.~(\ref{pairwise_potential}), represented here by the thick blue line. We also plot the canonical Lennard-Jones potential (thin line) for comparison. (b) During the preparation of our glassy samples we allow the effective size degrees of freedom $\lambda_i$ to change; their fluctuations are governed by the potential $\mu(\lambda_i,\lambda_i^{(0)})$ given by Eq.~(\ref{size_dof_potential}), and plotted here for various values of the stiffness $k_\lambda$ as indicated by the legend.}
\label{potential_fig}
\end{figure}

The effective sizes $\lambda_i$ are subjected to the potential
\begin{equation}\label{size_dof_potential}
\mu(\lambda_i, \lambda_i^{(0)}) = \sFrac{1}{2}k_\lambda(\lambda_i - \lambda_i^{(0)})^2\big(\sFrac{\lambda_i^{(0)}}{\lambda_i}\big)^2\,,
\end{equation}
where $\lambda_i^{(0)}$ is the energetically-favorable effective size of the $i\th$ particle in the absence of other interactions, and $k_\lambda$ is the stiffness associated with the effective size degrees of freedom (DOF). We will demonstrate in what follows that $k_\lambda$ plays a crucial role in determining the stability of our computer glasses. The potentials $\varphi(r_{ij},\lambda_i,\lambda_j)$ and $\mu(\lambda_i, \lambda_i^{(0)})$ are plotted in Fig.~\ref{potential_fig}. We
emphasize that in order to maintain a fixed equilibrium polydispersity, the potential $\mu$ should in general depend on temperature and pressure. (However, we expect the variations of $\mu$ to be small in the realistic setting of fixed pressure and varying temperature. In that case, fixing $\mu$ corresponds to a system of particles that slightly dilate with temperature, an effect which is unlikely to significantly affect  properties near the glass transition.)
By quenching at fixed $\mu$ from some temperature $T$ as we do, we generate inherent structures  characterizing the landscape at that temperature, structures that turn out to be ultra-stable.

We employ a 50:50 mixture, such that one half of the particles have $\lambda^{(0)}\!=\!0.5$, and the other half have $\lambda^{(0)}\!=\!0.7$, expressed in microscopic units of length $\ell$. All particles share the same mass $m$, and times are expressed in terms of $\sqrt{m\ell^2/\varepsilon}$. All physical observables presented in what follows should be understood as expressed in terms of the relevant microscopic units.

\subsection{Glass preparation protocol}

We created glassy samples as follows: we begin by fixing the number density $\rho = N/V$ (with $V$ the system's volume) at $0.5$, and performing a high-temperature ($T=1.0$) equilibration of the system subjected to the potential energy ${\cal U}$. For this part of the preparation protocol we choose the mass associated with the size DOF to be unity for all particles. We then employ the FIRE algorithm \cite{fire} to minimize the potential energy ${\cal U}$. This minimization is done while fixing the imposed pressure at $p\!=\!1.0$ using a Berendsen barostat \cite{berendsen}, with a time constant $\tau_{\mbox{\tiny Ber}}\!=\!10.0$. States are deemed to be in mechanical equilibrium when the ratio of the typical gradient of the potential to the typical interparticle force drops below $10^{-12}$. Crucially, upon convergence of the minimization of the potential energy ${\cal U}$, \emph{we freeze the effective size DOF for all subsequent analyses and simulations}, reducing the potential energy to
\begin{equation}\label{foo00}
U = \sum_{i<j}\varphi(r_{ij},\lambda_i,\lambda_j)\,,
\end{equation}
where the pairwise potential $\varphi$ given by Eq.~(\ref{pairwise_potential}) remains unchanged. Notice that, in contrast with ${\cal U}$, $U$ does not depend on the target effective sizes $\{\lambda_i^{(0)}\}$. By construction, configurations found by minimizing the potential ${\cal U}$ with respect to particle coordinates and effective sizes also correspond to local minima of the reduced potential $U$. 

We carried out the procedure explained above while systematically varying the stiffness $k_\lambda$ between $10$ and $10^5$. We have also created glasses in which the size DOF are completely frozen during glass formation, corresponding to the limit $k_\lambda\!\to\!\infty$. For the structural analyses and elasticity calculations discussed in the next section, we generated \numprint{42000} independent glassy samples of $N\!=\!\numprint{4000}$ particles for each value of the stiffness $k_\lambda$. For the shear-deformation experiments presented in Sect.~\ref{elasto_plastic_transients}, we generated a few tens of larger systems of $N\!=\!\numprint{256000}$ particles for values of $k_\lambda$ between $10$ and $10^4$, in addition to a hundred solids of $N\!=\!\numprint{16000}$ particles for values of $k_\lambda$ between $10^2$ and $10^5$.

\section{Structural analyses and elasticity}
\label{structure_and_elasticity}

\subsection{Structure}
\begin{figure}[!ht]
\centering
\includegraphics[width = 0.50\textwidth]{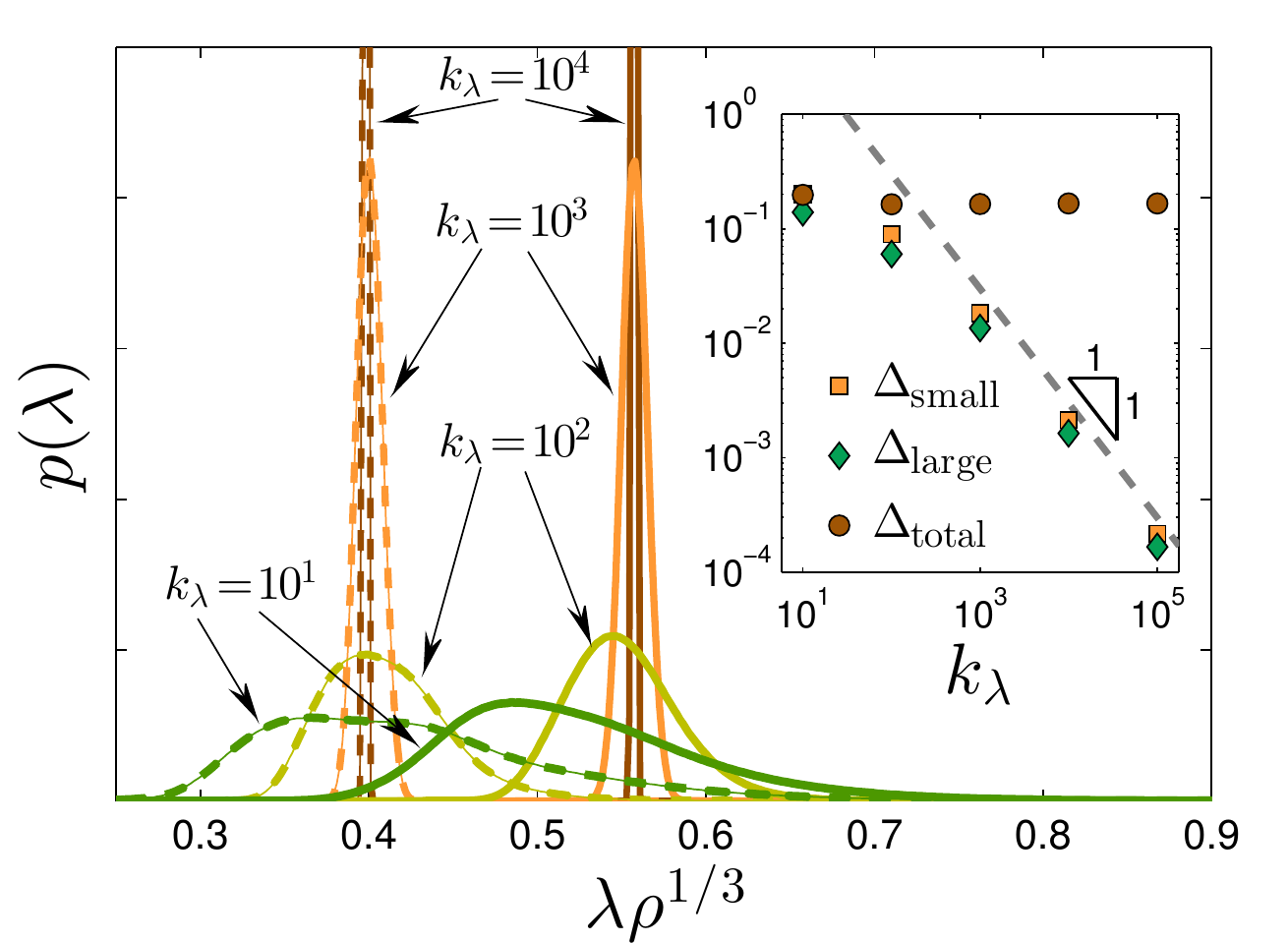}
\caption{\footnotesize Distributions $p(\lambda)$ of particles' effective size degrees of freedom, plotted against the dimensionless effective size $\lambda \rho^{1/3}$, for various values of $k_\lambda$ as indicated by the arrows. The dashed lines represent the distributions measured for particles whose target effective size during glass preparation was $\lambda^{(0)}\!=\!0.5$, and the solid lines represent those for which $\lambda^{(0)}\!=\!0.7$. 
Inset: the polydispersity $\Delta$ is defined as the ratio of the effective sizes' standard deviation to their mean. We report $\Delta$ for `small' and `large' particles, and the total polydispersity (calculated as the standard deviation to mean ratio, taken over all particles), vs.~the stiffnesses $k_\lambda$.}
\label{effective_size_distributions_fig}
\end{figure}

Our choice of chemical potential $\mu$ (parametrized by the stiffness $k_\lambda$, see Eq.~(\ref{size_dof_potential})) fixes the distribution of radii of the obtained inherent states, as we now characterize. Fig.~\ref{effective_size_distributions_fig} shows the distributions $p(\lambda)$ of effective particle sizes $\lambda$. As the stiffness $k_\lambda$ associated with the effective size DOF is reduced, the width of the distributions grows. In the large-$k_\lambda$ limit, we find $\Delta\!\sim\! k_\lambda^{-1}$, where $\Delta$ is the polydispersity (see figure caption for details), consistent with the prediction put forward in \cite{swap_jamming_prx_2018}. We find that below $k_\lambda\!\approx\!10^2$, the two peaks corresponding to `large' and `small' particles start to overlap. 

\begin{figure}[!ht]
\centering
\includegraphics[width = 0.50\textwidth]{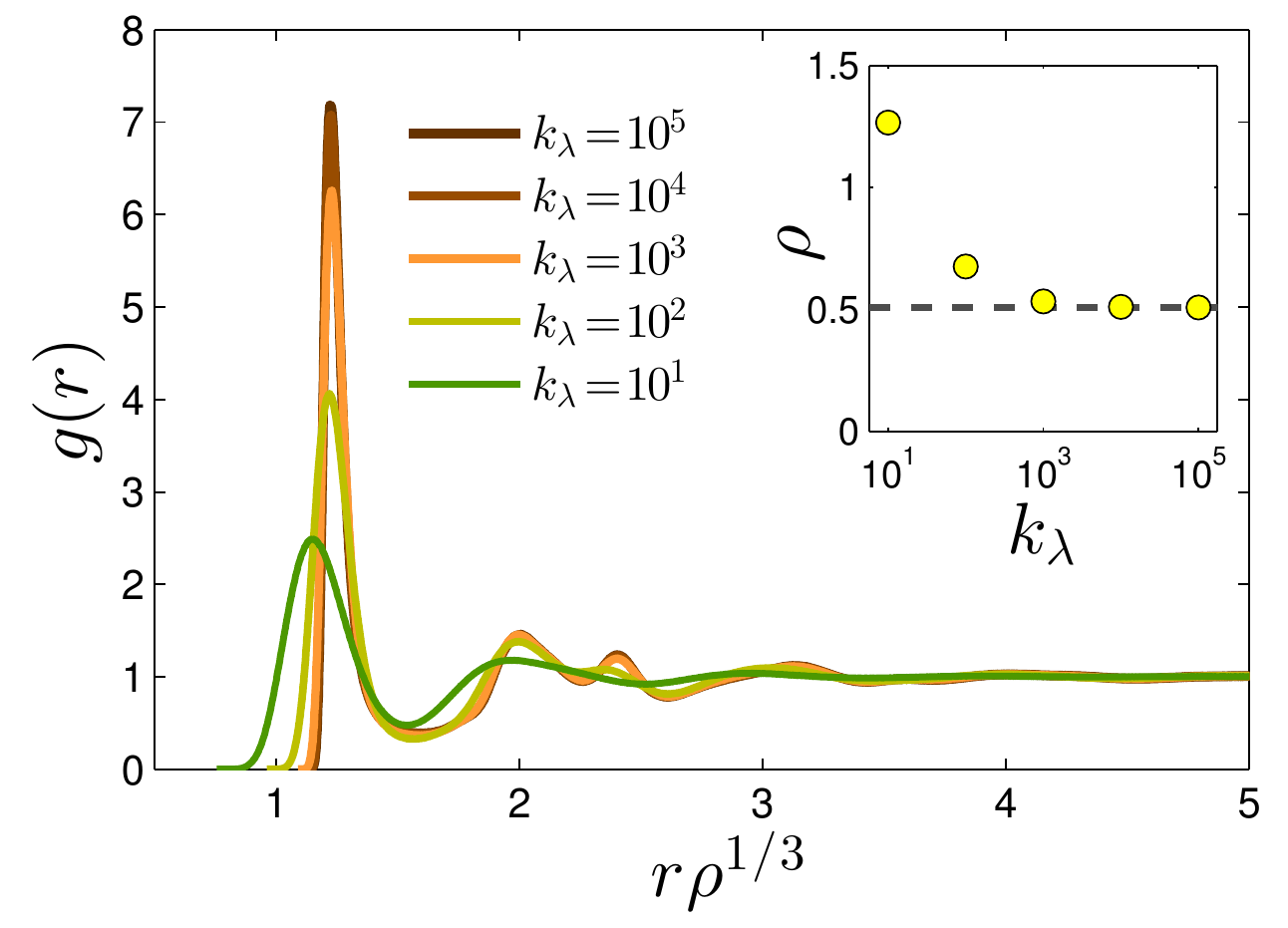}
\caption{\footnotesize Pair correlations $g(r)$ between pairs of `large' particles (i.e.~those with a target effective size $\lambda^{(0)}\!=\!0.70$ during glass preparation) measured in our glassy samples, plotted against the rescaled pairwise distances $r\rho^{1/3}$ for various values of the stiffness $k_\lambda$, increasing from thin to thick lines. We find no ordering upon decreasing $k_\lambda$. The inset shows how the density $\rho\!=\! N/V$ of our glassy samples increases as $k_\lambda$ is reduced. The horizontal dashed line marks the density of our $k_\lambda\!=\!\infty$ glasses. }
\label{pair_correlations}
\end{figure}

Next we study the pair correlation function $g(r)$ for various values of $k_{\lambda}$ in Fig.~\ref{pair_correlations}. As traditionally done, we calculated the pair correlations for pairs with the same `large' effective target size, the same `small' effective target size, and for different (`large'-`small') effective target sizes. In the figure we only present the `large'-`large' correlation function; the other two have similar features. We find that varying $k_\lambda$ does not seem to introduce any observable ordering. In fact, for smaller $k_\lambda$ the second and third peaks of $g(r)$ are diminished. We conclude that all of our constructed glassy samples are disordered.

\subsection{Macroscopic elasticity}
We next turn to examining the elastic properties of our glassy samples. We focus first on characterizing the degree of structural frustration that our glasses possess, as manifested by their sample-to-sample shear stress fluctuations. In particular, we are interested in assessing whether allowing for size fluctuations of the particles during glass formation reduces in some way the degree of structural frustration. In order to meaningfully compare between different system sizes and the ensembles created with different values of the size DOF stiffness $k_\lambda$, we calculate a system-size-independent and dimensionless sample-to-sample standard deviation of the shear stress by rescaling the dimensionful standard deviation $\delta\sigma$ by $1/\sqrt{N}$ and the athermal shear modulus $G$:
\begin{equation}
\delta \tilde{\sigma} \equiv \sqrt{N}\,\delta \sigma/G\,.
\end{equation}%
For the definition of $G$ and other elastic moduli, see Appendix~\ref{observables_appendix}. In Fig.~\ref{elasticity_fig}a we plot $\delta \tilde{\sigma}$ vs.~the stiffness $k_\lambda$.
As expected, lowering $k_{\lambda}$ results in more optimally packed glasses with a lower degree of structural frustration, as expressed by a decrease of a factor of two of the shear stress fluctuations over the entire range of $k_{\lambda}$.

\begin{figure}[!ht]
\centering
\includegraphics[width = 0.50\textwidth]{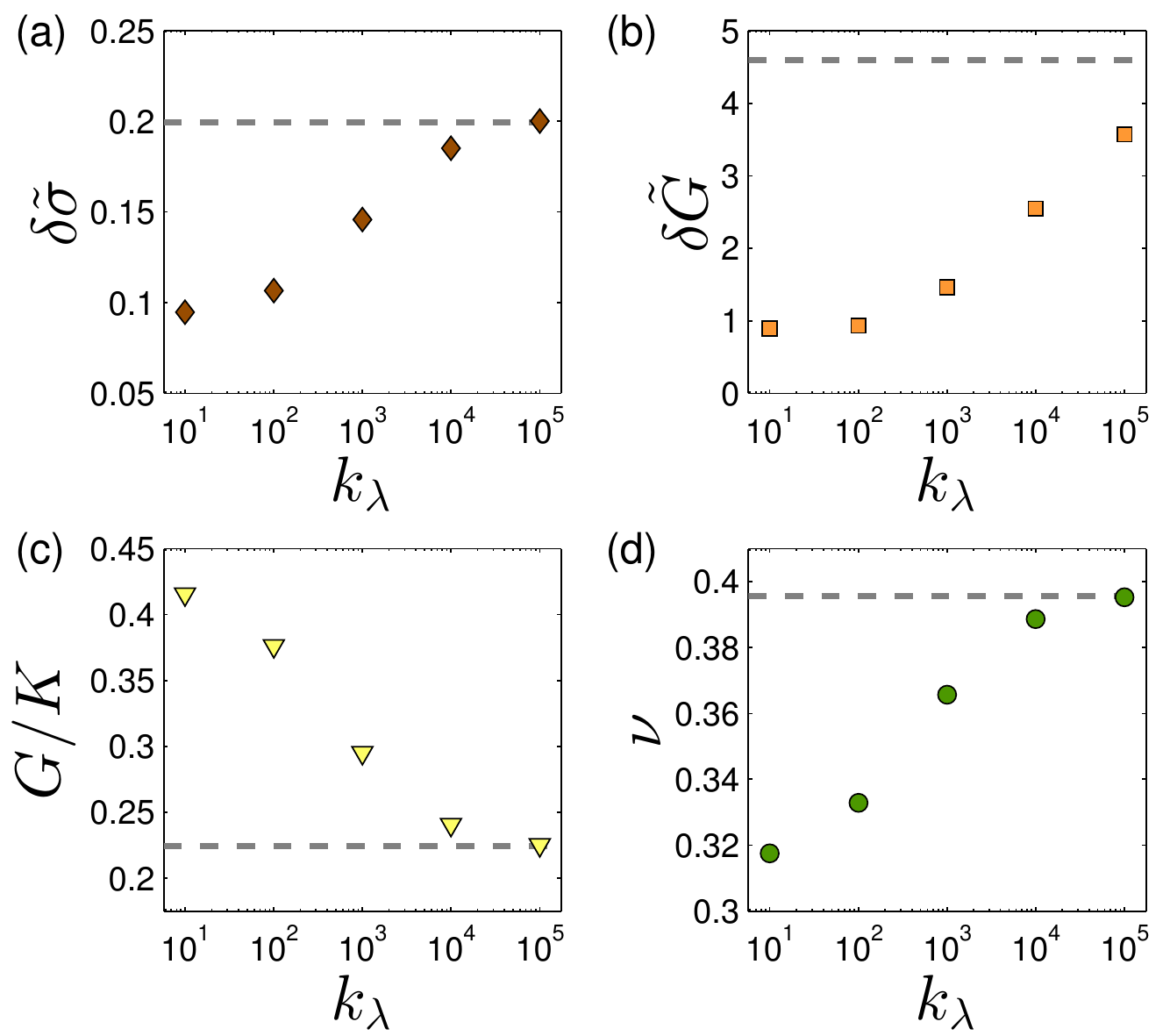}
\caption{\footnotesize Elastic properties of our glassy samples: (a) sample-to-sample standard deviation of the dimensionless shear stress $\delta\tilde{\sigma}$ (see text for definition and discussion), vs.~the stiffness of the size DOF $k_\lambda$. (b) $\delta\tilde{G}$ is the sample-to-sample standard deviation to mean ratio of the shear modulus, scaled by $\sqrt{N}$. (c) Sample-to-sample mean of the ratio of shear to bulk moduli. (d) Sample-to-sample mean Poisson's ratio of our glassy samples. The horizontal dashed lines represent the $k_\lambda\!=\!\infty$ values in all panels.}
\label{elasticity_fig}
\end{figure}

We next study the athermal elastic moduli of our glassy samples. In Fig.~\ref{elasticity_fig}b we report a dimensionless and system-size-independent characterizer of shear modulus fluctuations, defined as 
\begin{equation}
\delta\tilde{G} \equiv \sqrt{N}\,\delta G/G\,.
\end{equation}
Remarkably, the relative fluctuations decrease by over a factor of 4 across the entire sampled range of $k_\lambda$, suggesting that the increased stability of our glasses with decreasing $k_\lambda$ is accompanied by a strong reduction of coarse-grained local elastic moduli fields.

In Fig.~\ref{elasticity_fig}c we plot the sample-to-sample mean of our glasses' athermal shear to bulk moduli ratio (see definitions in Appendix~\ref{observables_appendix}). The ratio appears to increase above the $k_\lambda\!=\!\infty$ value --- represented by the horizontal dashed line --- by approximately $85\%$, which amounts to a variation of the glasses' Poisson's ratio from $\nu\!\approx\!0.4$ for $k_\lambda\!=\!\infty$ to $\nu\!\approx\!0.32$ for $k_\lambda\!=\!10$, as reported in Fig.~\ref{elasticity_fig}d. We emphasize that all of the above elastic properties show significant change over the range of measured $k_{\lambda}$, but start to saturate at around $k_{\lambda}\!=\!10^2$. We will see that this behaviour is consistent with our other measurements in the sections below.

\subsection{Vibrational spectra}
\label{dos}
The stability of disordered solids is often characterized in terms of the statistical properties of low-frequency vibrational modes that emerge due to the solids' disordered microstructure \cite{ohern2003, eric_boson_peak_emt, cge_paper, modes_ultra_stable_LB}. In particular, the destabilizing effect of internal stresses and structural frustration has been captured by Effective Medium \cite{eric_boson_peak_emt} and mean-field \cite{silvio} calculations, that predict a gapless spectrum $D(\omega)\!\sim\!\omega^2$ of non-phononic (i.e.~disorder-induced) vibrational modes. However, numerical results in spatial dimensions $\dbar\!\le\!4$ indicate that the non-Debye low-frequency spectrum (obtained by eliminating Goldstone modes, by considering small systems \cite{modes_prl,geert_non_debye} or by selecting modes based on their participation ratio \cite{ikeda_pnas,modes_ultra_stable_LB}) of generic structural computer glasses follows a universal $D(\omega)\!\sim\!\omega^4$ form, independent of model \cite{modes_prl}, preparation protocol \cite{cge_paper, modes_ultra_stable_LB}, proximity to the unjamming transition \cite{ikeda_pnas}, and spatial dimension \cite{geert_non_debye}. The modes that populate the $\omega^4$ tails have been shown to be quasilocalized \cite{modes_prl,geert_non_debye}, have been argued to control elasto-plastic responses of externally loaded-glasses \cite{plastic_modes_prerc,wencheng} and the singularities observed in nonlinear elastic moduli \cite{exist}. Furthermore, they are believed \cite{soft_potential_model_02,soft_potential_model_1991,Buchenau_1992,parisi_spin_glass} to serve as the tunneling two-level systems responsible for the universal anomalous thermodynamics of glasses below a few Kelvin \cite{pohl_1971,Anderson}.

\begin{figure}[!ht]
\centering
\includegraphics[width = 0.50\textwidth]{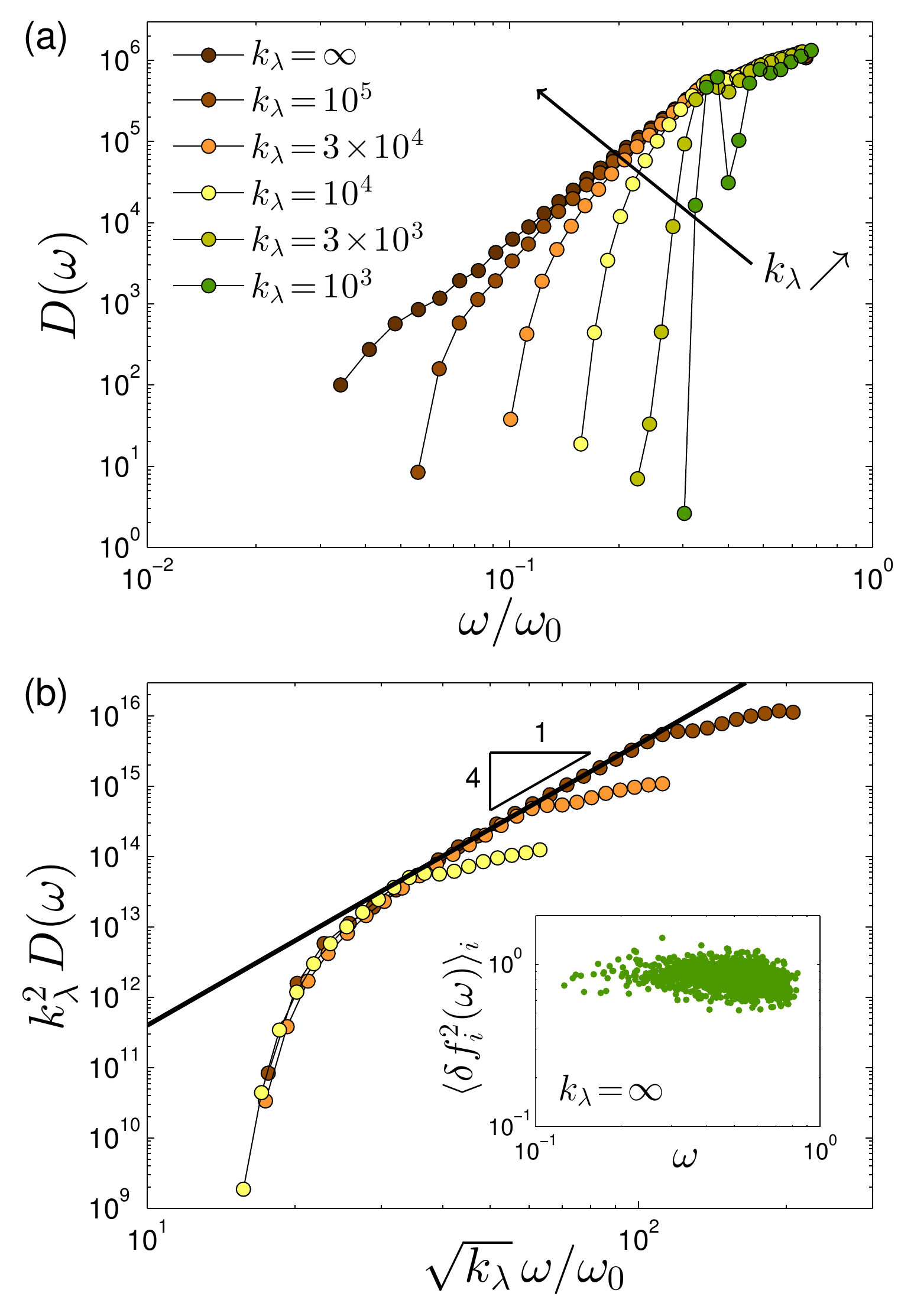}
\caption{\footnotesize (a) Density of vibrational modes $D(\omega)$ vs.~rescaled frequency $\omega/\omega_0$ where $\omega_0\!\equiv\! c_s/\rho^{-1/3}$ and $c_s\!\equiv\!\sqrt{G/\rho}$ is the shear wave speed. (b) In the limit of large $k_\lambda$ a gap of order $1/\sqrt{k_\lambda}$ opens in $D(\omega)$ as the collapse indicates. Here we plot $D(\omega)$ for $k_\lambda\!=\!10^5,3\!\times\!10^4$ and $10^4$. For smaller $k_{\lambda}$, the occurrence of the lowest-frequency phonons in $D(\omega)$ destroys the collapse. Inset: numerical validation of the frequency independence of the particle-wise mean squared variation of the forces $f_i\!\equiv\!-\partial U/\partial\lambda_i$, induced by a quasilocalized vibrational mode of frequency $\omega$ in $k_\lambda\!=\!\infty$ glasses, see discussion in Sect.~\ref{discussion}.}
\label{dos_fig}
\end{figure}

We study the statistics of non-phononic low-frequency vibrational modes of our glassy samples. We have calculated the Hessian matrix ${\cal M}\!\equiv\!\partial^2U/\partial\xv\partial\xv$ of each member of our ensembles of \numprint{42000} glassy solids, and calculated the first 120 vibrational modes (excluding the three translational zero modes). The resulting spectra are plotted in Fig.~\ref{dos_fig}a. We find that as $k_\lambda$ is decreased, a gap forms at low frequencies. This indicates that for our small-$k_\lambda$ glassy samples, quasilocalized modes are strongly depleted. 

In Fig.~\ref{dos_fig}b we show that the gap $\omega_{\mbox{\tiny$\Delta$}}\!\sim\!1/\sqrt{k_\lambda}\!\sim\!\sqrt{\Delta}$ in systems made with $k_\lambda\!\ge\!10^4$ by plotting $k_\lambda^2\,D(\omega)$ against $\sqrt{k_\lambda}\,\omega/\omega_0$, leading to a data collapse at low frequencies. We will explain the scaling of the gap with $k_\lambda$ in Sect.\ref{discussion}. For smaller $k_\lambda$, the occurrence of the lowest-frequency phonons in $D(\omega)$ destroys the collapse. We further find a consistent behavior with the previously observed $D(\omega)\!\sim\!\omega^4$ above the gap frequency scale.

\subsection{Nonaffine displacements}
Our measurement of the vibrational density of states in Sect.~\ref{dos} was limited to the range $k_\lambda\!\ge\!10^3$ since the lowest-frequency phonons hinder a clear observation of the further depletion of quasilocalized modes upon decreasing $k_\lambda$ beyond $k_{\lambda}\!=\!10^3$. We therefore supplement the measurements of the vibrational density of states with a study of the statistics of particles' linear displacement responses to forces that emerge following an imposed shear deformation, often referred to as the \emph{nonaffine} displacements, and denoted here by $\vv$. Nonaffine displacements are defined as
\begin{equation}
\vv = -{\cal M}^{-1}\cdot\frac{\partial^2U}{\partial\xv\partial\gamma}\,,
\end{equation}
where $\gamma$ parametrizes the imposed shear deformation, see details in Appendix.~\ref{aqs}. The main contributions to the contraction of ${\cal M}^{-1}$ with $\partial^2U/\partial\xv\partial\gamma$ are expected to stem from soft quasilocalized modes, rather than from low-frequency phonons \cite{exist}. The statistics of nonaffine displacements are therefore expected to echo those of soft quasilocalized vibrational modes. 

\begin{figure}[!ht]
\centering
\includegraphics[width = 0.50\textwidth]{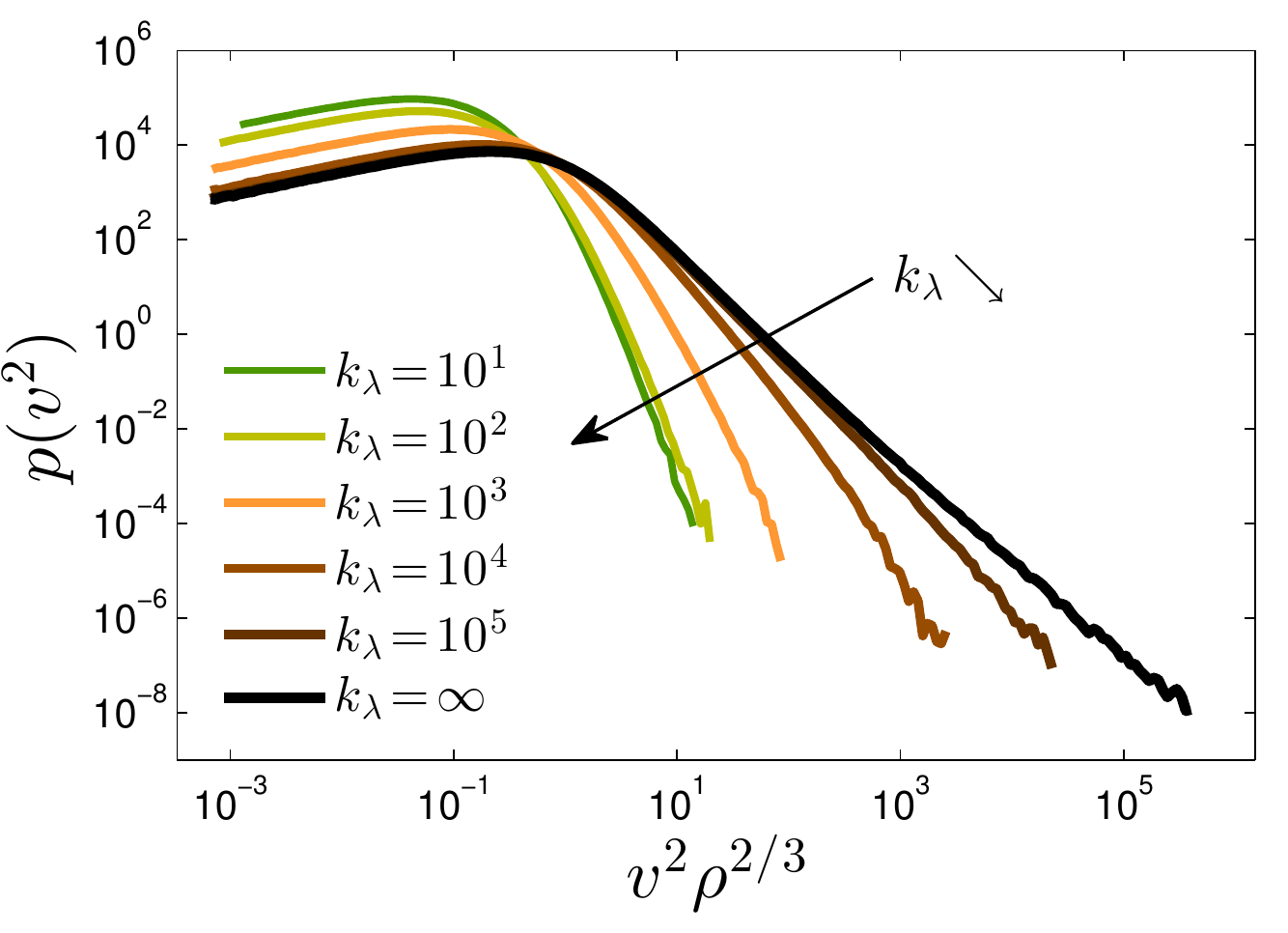}
\caption{\footnotesize Particle-wise distributions $p(v^2)$ of nonaffine displacements squared $v^2$, plotted against the dimensionless displacements squared $v^2\rho^{2/3}$, measured in our ensembles of glassy samples with various values of the size DOF $k_\lambda$, increasing from thin to thick lines. }
\label{nonaffine_fig}
\end{figure}

\begin{figure*}[!ht]
\centering
\includegraphics[width = 0.9\textwidth]{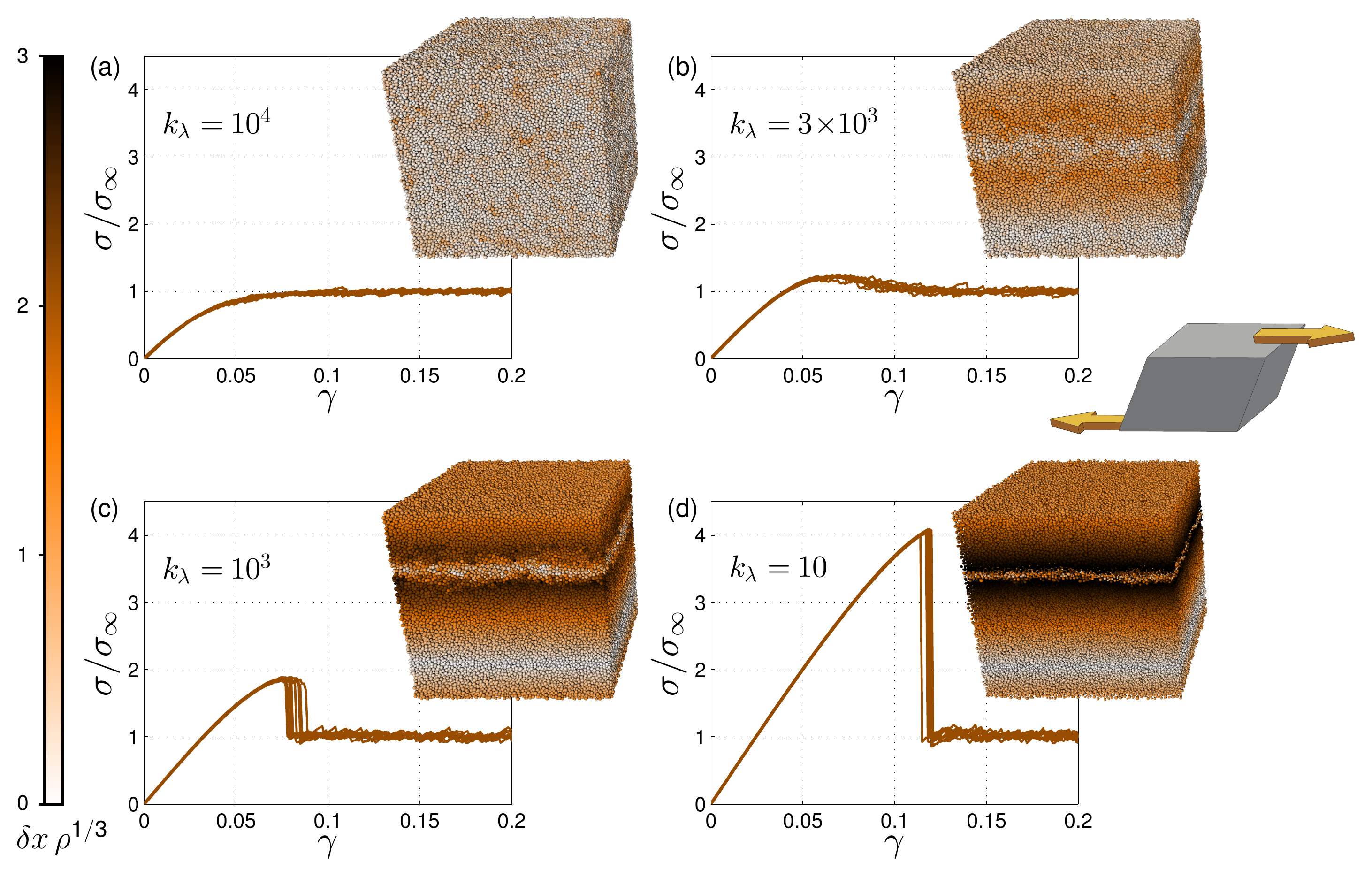}
\caption{\footnotesize Stress-strain curves for 
 computer glasses made with $k_\lambda\!=\!10^4,3\times10^3,10^3$ and $10^1$, from (a) to (d). Stresses are rescaled by their respective average steady-state values. The large insets show representative snapshots of the deformed solids; the color coding represents the magnitude of total nonaffine displacements $\delta x$ measured between $\gamma\!=\!0$ and $\gamma\!=\!0.13$, rescaled by the length $\rho^{-1/3}$. The small inset illustrates the geometry of the applied shear deformation.}
\label{aqs_fig}
\end{figure*}

In Fig.~\ref{nonaffine_fig} we show the particle-wise distributions of nonaffine displacements squared $v_i^2\!\equiv\!\vv_i\!\cdot\!\vv_i$ (no summation on $i$ implied). We indeed find that the form of the large-value tails of these distributions mirror the observed gaps in the density of vibrational modes as shown in Fig.~\ref{dos_fig}. Here, however, we are able to meaningfully probe the full range of $k_\lambda$ compared to the limited range shown in Fig.~\ref{dos_fig}. We further see the beginning of the saturation of the stabilizing effect below $k_\lambda\!=\!10^2$, consistent with the behavior of  the elastic properties reported in Fig.~\ref{elasticity_fig}.

There is an intimate relation between nonaffine displacements and the shear modulus: $G\!\propto\!\partial^2U/\partial\gamma^2\! + \!\vv\cdot\partial^2U/\partial\xv\partial\gamma$ \cite{lutsko}. The substantial reduction in the extreme values of the nonaffine displacements observed upon reducing $k_\lambda$ correlate with the decrease in sample-to-sample fluctuations of the shear modulus as seen in Fig.~\ref{elasticity_fig}b. We observe a saturation in both quantities for $k_\lambda\!\le\!10^2$.

\section{Elasto-plastic transients}
\label{elasto_plastic_transients}
In this Section we put the mechanical stability of our  ultrastable glasses to a direct test. We employ systems of \numprint{256000} particles and deform them under simple shear strain using an athermal quasistatic protocol as described in Appendix~\ref{aqs}. The results are presented in Fig.~\ref{aqs_fig} for glasses made with $k_\lambda\!=\!10^4,3\!\times\!10^3,10^3$ and $10^1$ in panels (a)-(d), respectively. Stress-strain curves for $k_\lambda\!=\!10^2$ are displayed in Fig.~\ref{shear_band_geometry_fig} of Appendix~\ref{aqs}. In these plots we rescaled the stress by its average steady-state value.

\begin{figure*}[!ht]
\centering
\includegraphics[width = 1.00\textwidth]{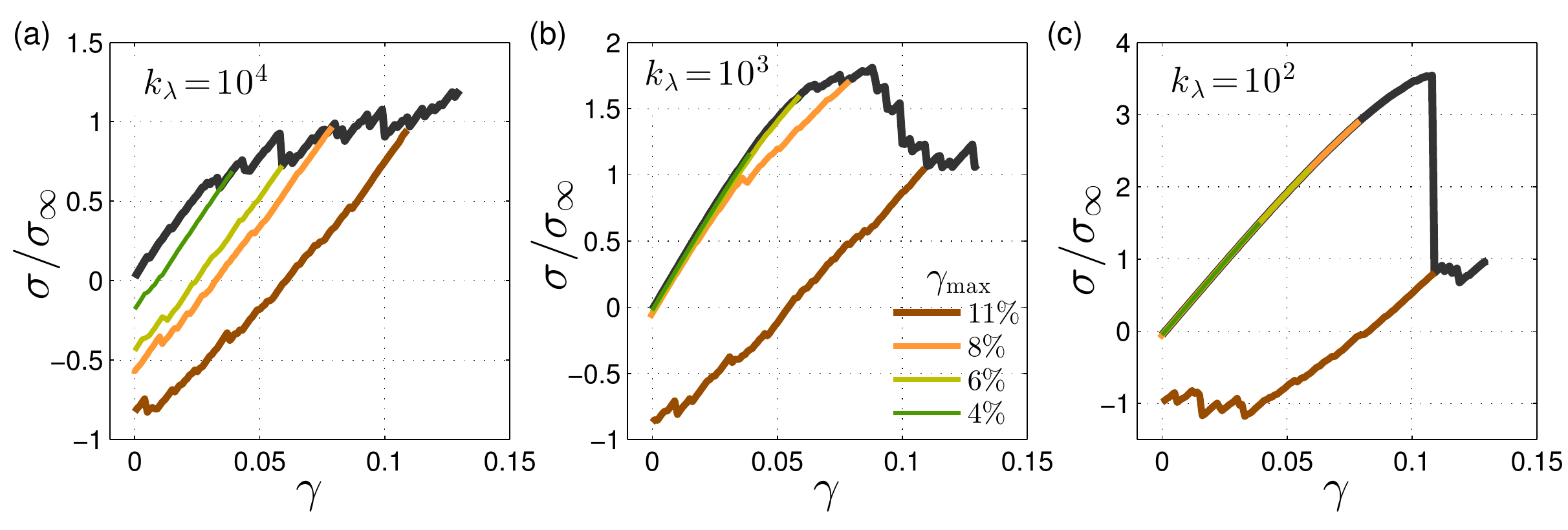}
\caption{\footnotesize Example stress-strain signals obtained in shear cycles, for $k_\lambda\!=\!10^4$ (a) $10^3$ (b) and $10^2$ (c), and for maximal strains $\gamma_{\mbox{\tiny max}}$ as indicated by the legend (growing from thin to thick). Curves are rescaled by the same steady-state stresses $\sigma_\infty$ as used in Fig.~\ref{aqs_fig}. For large $k_\lambda$ the dynamics is irreversible, even for small $\gamma_{\mbox{\tiny max}}$. For intermediate $k_\lambda$ some plasticity sets in before the macroscopic shear banding event; however the system is able to nearly return to the zero-stress state upon reversing the strain. For our most stable glasses we observe almost no plasticity before the macroscopic shear band (see also Fig.~\ref{dissipation_fig} below).}
\label{shear_loop_fig}
\end{figure*}

Glasses created with $k_\lambda\!=\!10^4$ show a monotonic increase of the stress as deformation proceeds. For $k_\lambda\!=\!3\times10^3$ there is a mild stress overshoot, and for $k_\lambda\!\le\!10^3$, we observe a stress overshoot terminated by the occurrence of a macroscopic stress drop signaling the nucleation of a system spanning shear band at a characteristic strain that increases with decreasing $k_\lambda$. We note that Fig.~\ref{aqs_fig} only presents data from samples whose shear band is parallel to the $x$-$z$ plane; when the shear band is parallel to the $x$-$y$ plane, the stress does not attain a plateau value after the shear band nucleation, which is an artefact of the geometry of the Lees-Edwards periodic boundary conditions employed \footnote{We thank Misaki Ozawa for pointing out this detail.}, as shown in Appendix~\ref{aqs}.

The stability of our glasses can be quantified by the relative magnitude of the stress drop, i.e.~the ratio between the height of the stress peak and the following steady-state stress. This ratio is zero for $k_\lambda\!=\!10^4$, and grows to $\approx\!4$ for $k_\lambda\!=\!10$. For comparison, the most stable configurations presented in \cite{yielding_LB_2018} that were created by Swap Monte Carlo feature a relative stress drop of $\approx\!3$, i.e. it is smaller by roughly 25\% compared to the relative stress drop found in our most stable glasses. This difference establishes that our glassy samples' mechanical stability is similar compared to that of the Swap-Monte-Carlo-generated glasses. We emphasize here that the computational bottleneck in this numerical experiment is the deformation simulation, which takes roughly an order of magnitude more computation time compared to the preparation of our glassy samples of \numprint{256000} particles. 
We further note that the relative magnitude of the stress drop across the shear-banding event increases the most dramatically between $k_\lambda\!=\!10^3$ and $k_\lambda\!=\!10^2$, and saturates upon decreasing $k_{\lambda}$ from $10^2$ to $10$, consistent with the trend we have observed for elastic properties (reported in Fig.~\ref{elasticity_fig}), indicating a possible relation between stability and elasticity.  

To assess the degree of plastic deformation occurring along the elasto-plastic transients, we have performed single shear cycles on systems of \numprint{16000} particles; we deformed our glasses using the same athermal quasistatic scheme (described in Appendix~\ref{aqs}), up to various maximal strain values $\gamma_{\mbox{\tiny max}}$, and back to zero strain, as shown in Fig.~\ref{shear_loop_fig}. Interestingly, we find that at intermediate $k_\lambda$ of $10^3$ and $3\!\times\!10^{2}$, plastic events take place before the occurrence of the macroscopic shear band; however, upon reversal of the strain, the system appears to nearly return to its original, undeformed zero-stress state, see for example Fig.~\ref{shear_loop_fig}b. This behavior has been termed `partial irreversibility' in \cite{Yoshino2018}, where similar findings for well-annealed hard sphere glasses were reported. Upon further decreasing $k_\lambda$ to $10^2$, very few plastic events take place before the macroscopic shear band occurs.

\begin{figure}[!ht]
\centering
\includegraphics[width = 0.50\textwidth]{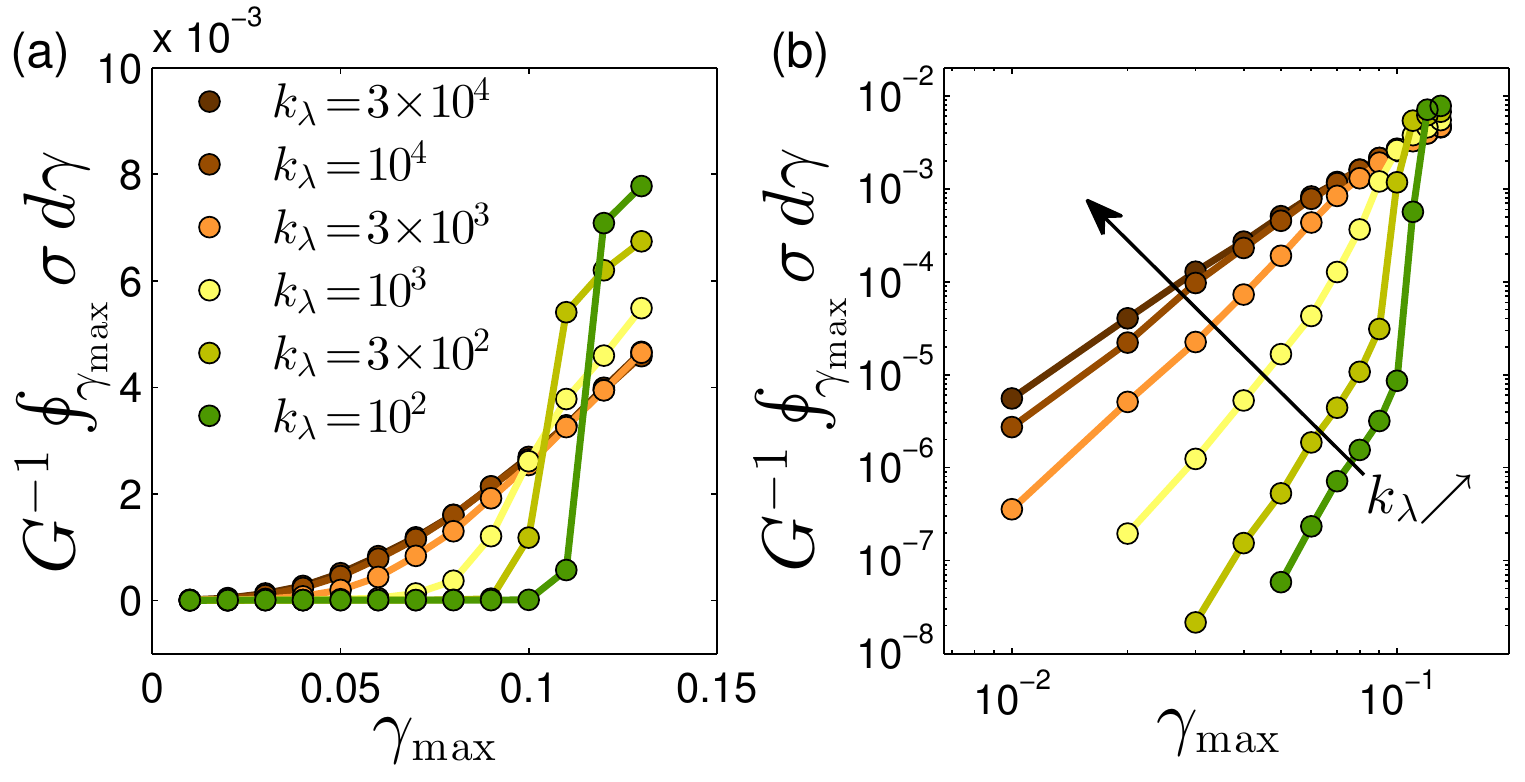}
\caption{\footnotesize Energy dissipated per unit volume (made dimensionless, see text) in a shear cycle of magnitude $\gamma_{\mbox{\tiny max}}$, reported in linear (a) and logarithmic (b) scales. }
\label{dissipation_fig}
\end{figure}

In Fig.~\ref{dissipation_fig} we report the sample-to-sample mean energy density dissipated in a shear cycle, made dimensionless by rescaling by the undeformed solids' shear modulus, namely $G^{-1}\oint_{\gamma_{\mbox{\fontsize{3}{0}\selectfont max}}}\sigma d\gamma$. Averages were taken over 100 independent realizations for each $k_\lambda$-ensemble. Consistent with the depletion of quasilocalized modes in the small $k_\lambda$ glasses, we observe a remarkably small degree of dissipation up to the macroscopic shear-banding event in those samples. 

\section{Scaling argument for the gap in the density of quasilocalized excitations}
\label{discussion}
Consider for a given $k_\lambda$ the lowest-frequency quasilocalized modes, which appear at a frequency scale $\omega_{\mbox{\tiny$\Delta$}}$. We shall argue that $\omega_{\mbox{\tiny$\Delta$}}\!\sim\!1/\sqrt{k_\lambda}$ or larger, otherwise the initial configuration would not be at a minimum of the energy function ${\cal  U}$ defined in Eq.~(\ref{111}). We denote by $\tilde{\cal M}$ the $N(\dbar\!+\!1)\!\times\! N(\dbar\!+\!1)$ hessian matrix of ${\cal U}$, which must be positive definite in any minimum. In the limit where $k_\lambda$ is infinite, the spectrum of $\tilde{\cal M}$ is the union of the spectrum of ${\cal M}$, together with additional modes at frequencies $\sim\!\sqrt{k_\lambda}$ corresponding to the ``breathing" of individual particles. For finite $k_\lambda$ the breathing modes hybridize with the usual vibrational modes, lowering the frequency of the latter. For large $k_\lambda$ this softening can be computed straightforwardly by perturbation theory \cite{swap_jamming_prx_2018}, and is of order $\Delta\omega^2\!\sim\! -\langle \delta f_i^2(\omega)\rangle_i /k_\lambda$ where $\delta f_i(\omega)\!=\!\sum_{j\ne i}\delta [\frac{\partial \varphi_{ij}}{\partial\lambda_{ij}}](\omega)$ is the compression induced by a mode of frequency $\omega$ on particle $i$, and $\langle\rangle_i$ indicates an average on all particles. In the inset of Fig.~\ref{dos_fig}b we scatter-plot $\langle \delta f_i^2(\omega)\rangle_i$ vs.~frequency $\omega$ of quasilocalized vibrational modes \footnote{We consider the lowest vibrational mode per glass in systems of $N\!=\!4000$ particles, which is typically quasilocalized \cite{modes_prl}} calculated in our $k_\lambda\!=\!\infty$ glasses, to find that they are independent of frequency.
 
Next we use the observation \cite{Ikeda2018,inst_note,Silbert2016PRE} that the low-frequency of quasilocalized modes stems from the near cancellation of two terms $\omega^2\!=\!\omega_+^2\!-\!\omega_-^2$. The contribution $\omega_+^2$ corresponds to the stretching of interactions whose characteristic stiffness is denoted $k$ and must scale as $\omega_+^2\!\sim\!\langle \delta f_i^2(\omega)\rangle_i /k$, whereas $\omega_-^2$ emerges due to pre-stress effects \cite{shlomo} and interactions with negative stiffnesses (usually absent in systems of purely repulsive particles). Counter-examples to this near cancellation can be found --- e.g.~rattlers in systems of purely repulsive particles. However (i) it can be shown to hold for modes causing the boson peak in a variety of systems \cite{matthieu_PRE_2005,matthieu_thesis,Xu07,eric_boson_peak_emt}, and (ii) the architecture of the modes forming the boson peak at frequency $\omega_{\mbox{\tiny BP}}$ was found to be essentially similar to that of quasilocalized modes, with $\omega_+^2\!\sim\!\omega_{\mbox{\tiny BP}}^2$~\cite{Ikeda2018}.

Using this result, we thus predict that $\omega_{\mbox{\tiny$\Delta$}}^2\!\geq\!\Delta\omega^2\!\sim\!\omega_{\mbox{\tiny BP}}^2 (k/k_\lambda)$, a bound indeed consistent with our observation. This bound, which must hold in all the inherent structures of ${\cal U}$, must also hold true for the ground state of the usual potential energy $U$. As a consequence, in continuously polydisperse materials, quasilocalized modes in very low-energy glassy configurations must be gapped.

If quasilocalized modes are gapped, then other excitations including shear transformations and two-level systems with small tunnelling barriers must also be gapped, since otherwise would imply the existence of vibrational modes in the forbidden frequency range \cite{wencheng}.  For example, using the fact that shear transformations sit near a saddle-node bifurcation, we expect  the characteristic stress at which plasticity sets in to scale as $\omega_{\mbox{\tiny$\Delta$}}^4\!\sim\!\Delta^2$.

\section{Summary and discussion}

In this work we introduced a simple computer glass former and preparation protocol --- following ideas put forward in \cite{swap_jamming_prx_2018} --- that enables the generation of  ultrastable glasses. By allowing the effective sizes of particles to fluctuate during glass formation, and freezing them thereafter, we are able to generate extremely stable glassy configurations at minimal computational cost. We demonstrated that the mechanical stability of our glasses is readily tunable by varying the stiffness $k_\lambda$ associated with the effective size DOF, and showed that it is at least comparable to the mechanical stability of glasses created using the Swap-Monte-Carlo method \cite{yielding_LB_2018}. Structural analyses reveal that no ordering takes place in any of our glasses.

Since our  ultrastable glasses are not created via a physical protocol, they may not be faithful representatives of real-world  glasses.
Also, their polydisperse nature, which is not a generic feature of structural glasses, is clearly key to their enhanced stability. This raises the crucial question of whether the structural and mechanical characteristics of our glasses are generic, or, conversely, that our glass-formation protocol introduces non-generic features.
 This resembles the open question posed in \cite{ultrastable_perez_pnas} of whether the absence of two-level systems in ultrastable vapor-deposited glasses is due to their increased stability (and hence, is a generic property of ultrastable glasses), or their preparation protocol.

The qualitative correspondence between the discontinuous response of bulk metallic glasses and that of our ultra-stable glasses is an encouraging item with regards to the genericity of our results. Our approach may thus help resolve which precise microstructural features of glassy solids are responsible for their mechanical stability. For example, results from our deformation simulations indicate that there should exist a critical stiffness $k_{\lambda,c}$ above which a discontinuous event nucleates, as predicted by several approaches \cite{Perez08,yielding_LB_2018,Popovic2018}. In \cite{Popovic2018} the anisotropy of the problem is included, and the discontinuous event for very stable glasses corresponds to a narrow shear band whose nucleation shares similarity with that of a fracture, a scenario that could be tested with our obtained configurations.


There is a qualitative difference between the nonphononic density of vibrational modes of glasses created with the Swap Monte Carlo algorithm and glasses created with our  approach. In \cite{modes_ultra_stable_LB} it was shown that glasses created by Swap Monte Carlo retain gapless non-Debye spectra, featuring $D(\omega)\!\sim\!\omega^4$ even for the most deeply annealed and stable glassy samples that can be created with that approach. This stands in contrast to the spectra of our  ultrastable glasses, that feature a gap for any finite $k_{\lambda}$. 

Two possible origins of the difference between these results are: (i)~glasses created with Swap Monte Carlo are quenched from a finite (although rather low) temperature, whereas our protocol produces glasses that undergo structural relaxation all the way down to zero temperature; (ii)~the equivalence between Swap Monte Carlo ultrastable glasses and our  ultrastable glasses is only expected in the thermodynamic limit \cite{Glandt_1984,Glandt_1987,swap_jamming_prx_2018}, in which a particle in a canonical (i.e.~with no external particle reservoir) Swap Monte Carlo system can assume any size within a finite support of relative width $\Delta$. However, in finite-size Swap Monte Carlo systems, a particle can only swap sizes with the $N\!-\!1$ other members of a single, particular realization of the polydispersity. This finite-size discretization, which is completely absent in our approach (our particles can assume any size), may introduce structural frustration and reduce the effectiveness of polydispersity in stabilizing the glass.

We note that the formation of a gap in the nonphononic density of vibrational modes also occurs when the degree of internal stresses is relieved by artificially reducing the magnitude of pairwise forces in model glasses \cite{eric_boson_peak_emt,inst_note}. Interestingly, measurements of a dimensionless characterization of sample-to-sample stress fluctuations in our glassy samples (see Fig.~\ref{elasticity_fig}a) also indicate a reduction of internal stresses with increasing stability. 


\acknowledgements
We thank Eric Lerner for his help with graphics, and M.~Popovic and T.~De Geus for discussions. E.~L.~acknowledges support from the Netherlands Organisation for Scientific Research (NWO) (Vidi grant no.~680-47-554/3259). M.~W.~thanks the Swiss National Science Foundation for support under Grant No. 200021-165509 and the Simons Foundation Grant ($\#$454953 Matthieu Wyart).

\appendix
\section{Observables}
\label{observables_appendix}

\subsection{Vibrational modes}
Vibrational modes were calculated by a numerical partial diagonalization of the dynamical matrix ${\cal M} \equiv \frac{\partial ^2 U}{\partial\xv\partial\xv}$, where $U$ is the potential energy as given by Eq.~(\ref{foo00}), and $\xv$ denotes the vector of $3N$ particles' Cartesian coordinates. We employed the ARPACK package \cite{arpack}.

\subsection{Athermal elastic moduli}
The shear stress is given by
\begin{equation}
  \sigma = \frac{1}{V}\frac{\partial U}{\partial \gamma}\,,
\end{equation}
where $\gamma$ denotes a simple shear strain. Athermal elastic moduli were calculated following the formulation of Lutsko \cite{lutsko}. The shear modulus $G$ is given by 
\begin{equation}
G = \frac{\frac{\partial^2U}{\partial\gamma^2} - \frac{\partial^2U}{\partial\gamma\partial\xv}\cdot{\cal M}^{-1}\cdot\frac{\partial^2U}{\partial\xv\partial\gamma}}{V}\,,
\end{equation}
and the bulk modulus $K$ by
\begin{equation}
K = \frac{\frac{\partial^2U}{\partial\eta^2} - \frac{\partial^2U}{\partial\eta\partial\xv}\cdot{\cal M}^{-1}\cdot\frac{\partial^2U}{\partial\xv\partial\eta}}{V\dbar^2} + p \,.
\end{equation}
Here $p\!\equiv\!-\frac{1}{V\dbar}\frac{\partial U}{\partial\eta}$ is the hydrostatic pressure, $V$ is the system's volume, and $\gamma,\eta$ are simple shear and expansive strains, respectively, that parametrize the 3D strain tensor as
\begin{equation}
\epsilon = \frac{1}{2}\left( \begin{array}{ccc}2\eta + \eta^2& \gamma  + \gamma\eta & 0\\ \gamma  + \gamma\eta & 2\eta + \eta^2 + \gamma^2 & 0 \\ 0 & 0 &2\eta + \eta^2 \end{array}\right)\,.
\end{equation}

\section{Athermal quasistatic deformation}
\label{aqs}
In addition to various static structural analyses of our glassy samples, we have also carried out conventional athermal quasistatic deformation simulations to test the stability of our glassy samples by studying their transient elasto-plastic response. We imposed increments of simple shear deformation by applying the following transformation of coordinates
\begin{eqnarray}
x_i &\to&  x_i + \delta\gamma\, y_i\,, \nonumber \\
y_i &\to& y_i \,, \nonumber \\
z_i &\to& z_i \,, 
\end{eqnarray}
using strain steps of $\delta\gamma\!=\!10^{-3}$. Each such transformation was followed by updating the images of the Lees-Edwards periodic boundary conditions \cite{allen1989computer}, and then minimizing the potential energy $U$ using a conventional conjugate gradient algorithm.

\begin{figure}[!ht]
\centering
\includegraphics[width = 0.50\textwidth]{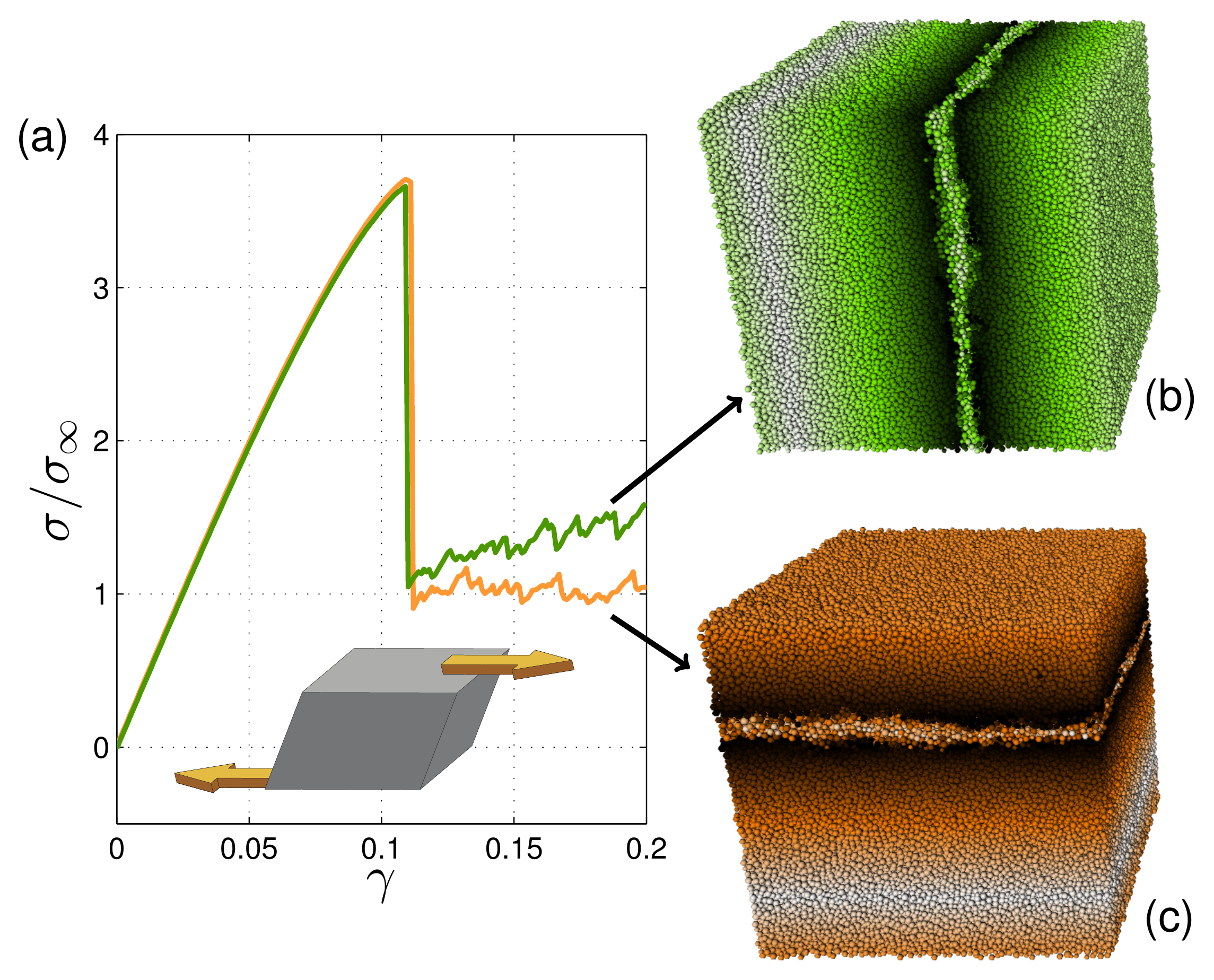}
\caption{\footnotesize (a) Stress-strain curves for  computer glasses made with $k_\lambda\!=\!10^2$, in which the two possible shear banding geometries occur, as shown in panels (b) and (c). The color code of particles, which represents the magnitude of nonaffine displacements, is the same as in Fig.~\ref{aqs_fig}.}
\label{shear_band_geometry_fig}
\end{figure}

In our deformation simulations, shear banding events can occur in two different geometries, as demonstrated in Fig.~\ref{shear_band_geometry_fig}: they can be parallel to the $x$-$z$ plane, as shown in panel (c), or parallel to the $y$-$z$ plane, as shown in panel (b). In Fig.~\ref{shear_band_geometry_fig}a we show that the resulting strain-strain curves in these two cases are different; in the former, the stress is stationary after the shear band (at least up to the maximal imposed deformation of 20\%), whereas in the latter the stress increases after the event. In Fig.~\ref{aqs_fig} we only show data pertaining to events with the geometry as displayed in panel (c).

\bibliography{references_lerner}

\begin{thebibliography}{49}%
\makeatletter
\providecommand \@ifxundefined [1]{%
 \@ifx{#1\undefined}
}%
\providecommand \@ifnum [1]{%
 \ifnum #1\expandafter \@firstoftwo
 \else \expandafter \@secondoftwo
 \fi
}%
\providecommand \@ifx [1]{%
 \ifx #1\expandafter \@firstoftwo
 \else \expandafter \@secondoftwo
 \fi
}%
\providecommand \natexlab [1]{#1}%
\providecommand \enquote  [1]{``#1''}%
\providecommand \bibnamefont  [1]{#1}%
\providecommand \bibfnamefont [1]{#1}%
\providecommand \citenamefont [1]{#1}%
\providecommand \href@noop [0]{\@secondoftwo}%
\providecommand \href [0]{\begingroup \@sanitize@url \@href}%
\providecommand \@href[1]{\@@startlink{#1}\@@href}%
\providecommand \@@href[1]{\endgroup#1\@@endlink}%
\providecommand \@sanitize@url [0]{\catcode `\\12\catcode `\$12\catcode
  `\&12\catcode `\#12\catcode `\^12\catcode `\_12\catcode `\%12\relax}%
\providecommand \@@startlink[1]{}%
\providecommand \@@endlink[0]{}%
\providecommand \url  [0]{\begingroup\@sanitize@url \@url }%
\providecommand \@url [1]{\endgroup\@href {#1}{\urlprefix }}%
\providecommand \urlprefix  [0]{URL }%
\providecommand \Eprint [0]{\href }%
\providecommand \doibase [0]{http://dx.doi.org/}%
\providecommand \selectlanguage [0]{\@gobble}%
\providecommand \bibinfo  [0]{\@secondoftwo}%
\providecommand \bibfield  [0]{\@secondoftwo}%
\providecommand \translation [1]{[#1]}%
\providecommand \BibitemOpen [0]{}%
\providecommand \bibitemStop [0]{}%
\providecommand \bibitemNoStop [0]{.\EOS\space}%
\providecommand \EOS [0]{\spacefactor3000\relax}%
\providecommand \BibitemShut  [1]{\csname bibitem#1\endcsname}%
\let\auto@bib@innerbib\@empty
\bibitem [{\citenamefont {Ketkaew}\ \emph {et~al.}(2018)\citenamefont {Ketkaew}
  \emph {et~al.}}]{fracture_toughness_eran_2018}%
  \BibitemOpen
  \bibfield  {author} {\bibinfo {author} {\bibfnamefont {J.}~\bibnamefont
  {Ketkaew}} \emph {et~al.},\ }\href {\doibase 10.1038/s41467-018-05682-8}
  {\bibfield  {journal} {\bibinfo  {journal} {Nat Commun.}\ }\textbf {\bibinfo
  {volume} {9}},\ \bibinfo {pages} {3271} (\bibinfo {year} {2018})}\BibitemShut
  {NoStop}%
\bibitem [{\citenamefont {Garrett}\ \emph {et~al.}(2016)\citenamefont
  {Garrett}, \citenamefont {Demetriou}, \citenamefont {Launey},\ and\
  \citenamefont {Johnson}}]{Garrett10257}%
  \BibitemOpen
  \bibfield  {author} {\bibinfo {author} {\bibfnamefont {G.~R.}\ \bibnamefont
  {Garrett}}, \bibinfo {author} {\bibfnamefont {M.~D.}\ \bibnamefont
  {Demetriou}}, \bibinfo {author} {\bibfnamefont {M.~E.}\ \bibnamefont
  {Launey}}, \ and\ \bibinfo {author} {\bibfnamefont {W.~L.}\ \bibnamefont
  {Johnson}},\ }\href {\doibase 10.1073/pnas.1610920113} {\bibfield  {journal}
  {\bibinfo  {journal} {Proc. Natl. Acad. Sci. U.S.A.}\ }\textbf {\bibinfo
  {volume} {113}},\ \bibinfo {pages} {10257} (\bibinfo {year}
  {2016})}\BibitemShut {NoStop}%
\bibitem [{\citenamefont {Shi}\ and\ \citenamefont {Falk}(2007)}]{SHI20074317}%
  \BibitemOpen
  \bibfield  {author} {\bibinfo {author} {\bibfnamefont {Y.}~\bibnamefont
  {Shi}}\ and\ \bibinfo {author} {\bibfnamefont {M.~L.}\ \bibnamefont {Falk}},\
  }\href {\doibase http://dx.doi.org/10.1016/j.actamat.2007.03.029} {\bibfield
  {journal} {\bibinfo  {journal} {Acta Mater.}\ }\textbf {\bibinfo {volume}
  {55}},\ \bibinfo {pages} {4317 } (\bibinfo {year} {2007})}\BibitemShut
  {NoStop}%
\bibitem [{\citenamefont {Lerner}\ and\ \citenamefont
  {Bouchbinder}(2018{\natexlab{a}})}]{cge_paper}%
  \BibitemOpen
  \bibfield  {author} {\bibinfo {author} {\bibfnamefont {E.}~\bibnamefont
  {Lerner}}\ and\ \bibinfo {author} {\bibfnamefont {E.}~\bibnamefont
  {Bouchbinder}},\ }\href {\doibase 10.1063/1.5024776} {\bibfield  {journal}
  {\bibinfo  {journal} {J. Chem. Phys.}\ }\textbf {\bibinfo {volume} {148}},\
  \bibinfo {pages} {214502} (\bibinfo {year} {2018}{\natexlab{a}})}\BibitemShut
  {NoStop}%
\bibitem [{\citenamefont {Lerner}\ and\ \citenamefont
  {Bouchbinder}(2018{\natexlab{b}})}]{inst_note}%
  \BibitemOpen
  \bibfield  {author} {\bibinfo {author} {\bibfnamefont {E.}~\bibnamefont
  {Lerner}}\ and\ \bibinfo {author} {\bibfnamefont {E.}~\bibnamefont
  {Bouchbinder}},\ }\href {\doibase 10.1103/PhysRevE.97.032140} {\bibfield
  {journal} {\bibinfo  {journal} {Phys. Rev. E}\ }\textbf {\bibinfo {volume}
  {97}},\ \bibinfo {pages} {032140} (\bibinfo {year}
  {2018}{\natexlab{b}})}\BibitemShut {NoStop}%
\bibitem [{\citenamefont {P\'erez-Casta\~{n}eda}\ \emph
  {et~al.}(2014)\citenamefont {P\'erez-Casta\~{n}eda}, \citenamefont
  {Rodr\'iguez-Tinoco}, \citenamefont {Rodr\'iguez-Viejo},\ and\ \citenamefont
  {Ramos}}]{ultrastable_perez_pnas}%
  \BibitemOpen
  \bibfield  {author} {\bibinfo {author} {\bibfnamefont {T.}~\bibnamefont
  {P\'erez-Casta\~{n}eda}}, \bibinfo {author} {\bibfnamefont {C.}~\bibnamefont
  {Rodr\'iguez-Tinoco}}, \bibinfo {author} {\bibfnamefont {J.}~\bibnamefont
  {Rodr\'iguez-Viejo}}, \ and\ \bibinfo {author} {\bibfnamefont {M.~A.}\
  \bibnamefont {Ramos}},\ }\href {\doibase 10.1073/pnas.1405545111} {\bibfield
  {journal} {\bibinfo  {journal} {Proc. Natl. Acad. Sci. U.S.A.}\ }\textbf
  {\bibinfo {volume} {111}},\ \bibinfo {pages} {11275} (\bibinfo {year}
  {2014})}\BibitemShut {NoStop}%
\bibitem [{\citenamefont {Ediger}(2017)}]{ediger_2017}%
  \BibitemOpen
  \bibfield  {author} {\bibinfo {author} {\bibfnamefont {M.~D.}\ \bibnamefont
  {Ediger}},\ }\href {\doibase 10.1063/1.5006265} {\bibfield  {journal}
  {\bibinfo  {journal} {J. Chem. Phys.}\ }\textbf {\bibinfo {volume} {147}},\
  \bibinfo {pages} {210901} (\bibinfo {year} {2017})}\BibitemShut {NoStop}%
\bibitem [{\citenamefont {Zeller}\ and\ \citenamefont
  {Pohl}(1971)}]{pohl_1971}%
  \BibitemOpen
  \bibfield  {author} {\bibinfo {author} {\bibfnamefont {R.~C.}\ \bibnamefont
  {Zeller}}\ and\ \bibinfo {author} {\bibfnamefont {R.~O.}\ \bibnamefont
  {Pohl}},\ }\href {\doibase 10.1103/PhysRevB.4.2029} {\bibfield  {journal}
  {\bibinfo  {journal} {Phys. Rev. B}\ }\textbf {\bibinfo {volume} {4}},\
  \bibinfo {pages} {2029} (\bibinfo {year} {1971})}\BibitemShut {NoStop}%
\bibitem [{\citenamefont {Anderson}\ \emph {et~al.}(1972)\citenamefont
  {Anderson}, \citenamefont {Halperin},\ and\ \citenamefont
  {Varma}}]{Anderson}%
  \BibitemOpen
  \bibfield  {author} {\bibinfo {author} {\bibfnamefont {P.~W.}\ \bibnamefont
  {Anderson}}, \bibinfo {author} {\bibfnamefont {B.~I.}\ \bibnamefont
  {Halperin}}, \ and\ \bibinfo {author} {\bibfnamefont {C.~M.}\ \bibnamefont
  {Varma}},\ }\href {\doibase 10.1080/14786437208229210} {\bibfield  {journal}
  {\bibinfo  {journal} {Philos. Mag.}\ }\textbf {\bibinfo {volume} {25}},\
  \bibinfo {pages} {1} (\bibinfo {year} {1972})}\BibitemShut {NoStop}%
\bibitem [{\citenamefont {Ninarello}\ \emph {et~al.}(2017)\citenamefont
  {Ninarello}, \citenamefont {Berthier},\ and\ \citenamefont
  {Coslovich}}]{berthier_prx}%
  \BibitemOpen
  \bibfield  {author} {\bibinfo {author} {\bibfnamefont {A.}~\bibnamefont
  {Ninarello}}, \bibinfo {author} {\bibfnamefont {L.}~\bibnamefont {Berthier}},
  \ and\ \bibinfo {author} {\bibfnamefont {D.}~\bibnamefont {Coslovich}},\
  }\href {\doibase 10.1103/PhysRevX.7.021039} {\bibfield  {journal} {\bibinfo
  {journal} {Phys. Rev. X}\ }\textbf {\bibinfo {volume} {7}},\ \bibinfo {pages}
  {021039} (\bibinfo {year} {2017})}\BibitemShut {NoStop}%
\bibitem [{\citenamefont {Tsai}\ \emph {et~al.}(1978)\citenamefont {Tsai},
  \citenamefont {Abraham},\ and\ \citenamefont {Pound}}]{TSAI1978465}%
  \BibitemOpen
  \bibfield  {author} {\bibinfo {author} {\bibfnamefont {N.-H.}\ \bibnamefont
  {Tsai}}, \bibinfo {author} {\bibfnamefont {F.~F.}\ \bibnamefont {Abraham}}, \
  and\ \bibinfo {author} {\bibfnamefont {G.}~\bibnamefont {Pound}},\ }\href
  {\doibase https://doi.org/10.1016/0039-6028(78)90134-6} {\bibfield  {journal}
  {\bibinfo  {journal} {Surf. Sci.}\ }\textbf {\bibinfo {volume} {77}},\
  \bibinfo {pages} {465 } (\bibinfo {year} {1978})}\BibitemShut {NoStop}%
\bibitem [{\citenamefont {Gazzillo}\ and\ \citenamefont
  {Pastore}(1989)}]{GAZZILLO1989388}%
  \BibitemOpen
  \bibfield  {author} {\bibinfo {author} {\bibfnamefont {D.}~\bibnamefont
  {Gazzillo}}\ and\ \bibinfo {author} {\bibfnamefont {G.}~\bibnamefont
  {Pastore}},\ }\href {\doibase https://doi.org/10.1016/0009-2614(89)87505-0}
  {\bibfield  {journal} {\bibinfo  {journal} {Chem. Phys. Lett.}\ }\textbf
  {\bibinfo {volume} {159}},\ \bibinfo {pages} {388 } (\bibinfo {year}
  {1989})}\BibitemShut {NoStop}%
\bibitem [{\citenamefont {Grigera}\ and\ \citenamefont
  {Parisi}(2001)}]{Grigera2001}%
  \BibitemOpen
  \bibfield  {author} {\bibinfo {author} {\bibfnamefont {T.~S.}\ \bibnamefont
  {Grigera}}\ and\ \bibinfo {author} {\bibfnamefont {G.}~\bibnamefont
  {Parisi}},\ }\href {\doibase 10.1103/PhysRevE.63.045102} {\bibfield
  {journal} {\bibinfo  {journal} {Phys. Rev. E}\ }\textbf {\bibinfo {volume}
  {63}},\ \bibinfo {pages} {045102} (\bibinfo {year} {2001})}\BibitemShut
  {NoStop}%
\bibitem [{\citenamefont {Guti\'{e}rrez}\ \emph {et~al.}(2015)\citenamefont
  {Guti\'{e}rrez}, \citenamefont {Karmakar}, \citenamefont {Pollack},\ and\
  \citenamefont {Procaccia}}]{smarajit_epl_2015}%
  \BibitemOpen
  \bibfield  {author} {\bibinfo {author} {\bibfnamefont {R.}~\bibnamefont
  {Guti\'{e}rrez}}, \bibinfo {author} {\bibfnamefont {S.}~\bibnamefont
  {Karmakar}}, \bibinfo {author} {\bibfnamefont {Y.~G.}\ \bibnamefont
  {Pollack}}, \ and\ \bibinfo {author} {\bibfnamefont {I.}~\bibnamefont
  {Procaccia}},\ }\href {http://stacks.iop.org/0295-5075/111/i=5/a=56009}
  {\bibfield  {journal} {\bibinfo  {journal} {Europhys. Lett.}\ }\textbf
  {\bibinfo {volume} {111}},\ \bibinfo {pages} {56009} (\bibinfo {year}
  {2015})}\BibitemShut {NoStop}%
\bibitem [{\citenamefont {Ozawa}\ \emph {et~al.}(2018)\citenamefont {Ozawa},
  \citenamefont {Berthier}, \citenamefont {Biroli}, \citenamefont {Rosso},\
  and\ \citenamefont {Tarjus}}]{yielding_LB_2018}%
  \BibitemOpen
  \bibfield  {author} {\bibinfo {author} {\bibfnamefont {M.}~\bibnamefont
  {Ozawa}}, \bibinfo {author} {\bibfnamefont {L.}~\bibnamefont {Berthier}},
  \bibinfo {author} {\bibfnamefont {G.}~\bibnamefont {Biroli}}, \bibinfo
  {author} {\bibfnamefont {A.}~\bibnamefont {Rosso}}, \ and\ \bibinfo {author}
  {\bibfnamefont {G.}~\bibnamefont {Tarjus}},\ }\href {\doibase
  10.1073/pnas.1806156115} {\bibfield  {journal} {\bibinfo  {journal} {Proc.
  Natl. Acad. Sci. U.S.A.}\ }\textbf {\bibinfo {volume} {115}},\ \bibinfo
  {pages} {6656} (\bibinfo {year} {2018})}\BibitemShut {NoStop}%
\bibitem [{\citenamefont {Lerner}\ \emph {et~al.}(2016)\citenamefont {Lerner},
  \citenamefont {D\"uring},\ and\ \citenamefont {Bouchbinder}}]{modes_prl}%
  \BibitemOpen
  \bibfield  {author} {\bibinfo {author} {\bibfnamefont {E.}~\bibnamefont
  {Lerner}}, \bibinfo {author} {\bibfnamefont {G.}~\bibnamefont {D\"uring}}, \
  and\ \bibinfo {author} {\bibfnamefont {E.}~\bibnamefont {Bouchbinder}},\
  }\href {\doibase 10.1103/PhysRevLett.117.035501} {\bibfield  {journal}
  {\bibinfo  {journal} {Phys. Rev. Lett.}\ }\textbf {\bibinfo {volume} {117}},\
  \bibinfo {pages} {035501} (\bibinfo {year} {2016})}\BibitemShut {NoStop}%
\bibitem [{\citenamefont {Kapteijns}\ \emph {et~al.}(2018)\citenamefont
  {Kapteijns}, \citenamefont {Bouchbinder},\ and\ \citenamefont
  {Lerner}}]{geert_non_debye}%
  \BibitemOpen
  \bibfield  {author} {\bibinfo {author} {\bibfnamefont {G.}~\bibnamefont
  {Kapteijns}}, \bibinfo {author} {\bibfnamefont {E.}~\bibnamefont
  {Bouchbinder}}, \ and\ \bibinfo {author} {\bibfnamefont {E.}~\bibnamefont
  {Lerner}},\ }\href {\doibase 10.1103/PhysRevLett.121.055501} {\bibfield
  {journal} {\bibinfo  {journal} {Phys. Rev. Lett.}\ }\textbf {\bibinfo
  {volume} {121}},\ \bibinfo {pages} {055501} (\bibinfo {year}
  {2018})}\BibitemShut {NoStop}%
\bibitem [{\citenamefont {Wang}\ \emph {et~al.}(2018)\citenamefont {Wang},
  \citenamefont {Ninarello}, \citenamefont {Guan}, \citenamefont {Berthier},
  \citenamefont {Szamel},\ and\ \citenamefont
  {Flenner}}]{modes_ultra_stable_LB}%
  \BibitemOpen
  \bibfield  {author} {\bibinfo {author} {\bibfnamefont {L.}~\bibnamefont
  {Wang}}, \bibinfo {author} {\bibfnamefont {A.}~\bibnamefont {Ninarello}},
  \bibinfo {author} {\bibfnamefont {P.}~\bibnamefont {Guan}}, \bibinfo {author}
  {\bibfnamefont {L.}~\bibnamefont {Berthier}}, \bibinfo {author}
  {\bibfnamefont {G.}~\bibnamefont {Szamel}}, \ and\ \bibinfo {author}
  {\bibfnamefont {E.}~\bibnamefont {Flenner}},\ }\href
  {https://arxiv.org/abs/1804.08765} {\bibfield  {journal} {\bibinfo  {journal}
  {arXiv preprint arXiv:1804.08765}\ } (\bibinfo {year} {2018})}\BibitemShut
  {NoStop}%
\bibitem [{\citenamefont {Brito}\ \emph {et~al.}(2018)\citenamefont {Brito},
  \citenamefont {Lerner},\ and\ \citenamefont {Wyart}}]{swap_jamming_prx_2018}%
  \BibitemOpen
  \bibfield  {author} {\bibinfo {author} {\bibfnamefont {C.}~\bibnamefont
  {Brito}}, \bibinfo {author} {\bibfnamefont {E.}~\bibnamefont {Lerner}}, \
  and\ \bibinfo {author} {\bibfnamefont {M.}~\bibnamefont {Wyart}},\ }\href
  {\doibase 10.1103/PhysRevX.8.031050} {\bibfield  {journal} {\bibinfo
  {journal} {Phys. Rev. X}\ }\textbf {\bibinfo {volume} {8}},\ \bibinfo {pages}
  {031050} (\bibinfo {year} {2018})}\BibitemShut {NoStop}%
\bibitem [{\citenamefont {Briano}\ and\ \citenamefont
  {Glandt}(1984)}]{Glandt_1984}%
  \BibitemOpen
  \bibfield  {author} {\bibinfo {author} {\bibfnamefont {J.~G.}\ \bibnamefont
  {Briano}}\ and\ \bibinfo {author} {\bibfnamefont {E.~D.}\ \bibnamefont
  {Glandt}},\ }\href {\doibase 10.1063/1.447087} {\bibfield  {journal}
  {\bibinfo  {journal} {J. Chem. Phys.}\ }\textbf {\bibinfo {volume} {80}},\
  \bibinfo {pages} {3336} (\bibinfo {year} {1984})}\BibitemShut {NoStop}%
\bibitem [{\citenamefont {Kofke}\ and\ \citenamefont
  {Glandt}(1987)}]{Glandt_1987}%
  \BibitemOpen
  \bibfield  {author} {\bibinfo {author} {\bibfnamefont {D.~A.}\ \bibnamefont
  {Kofke}}\ and\ \bibinfo {author} {\bibfnamefont {E.~D.}\ \bibnamefont
  {Glandt}},\ }\href {\doibase 10.1063/1.452800} {\bibfield  {journal}
  {\bibinfo  {journal} {J. Chem. Phys.}\ }\textbf {\bibinfo {volume} {87}},\
  \bibinfo {pages} {4881} (\bibinfo {year} {1987})}\BibitemShut {NoStop}%
\bibitem [{\citenamefont {Kob}\ and\ \citenamefont {Andersen}(1995)}]{kablj}%
  \BibitemOpen
  \bibfield  {author} {\bibinfo {author} {\bibfnamefont {W.}~\bibnamefont
  {Kob}}\ and\ \bibinfo {author} {\bibfnamefont {H.~C.}\ \bibnamefont
  {Andersen}},\ }\href {\doibase 10.1103/PhysRevE.51.4626} {\bibfield
  {journal} {\bibinfo  {journal} {Phys. Rev. E}\ }\textbf {\bibinfo {volume}
  {51}},\ \bibinfo {pages} {4626} (\bibinfo {year} {1995})}\BibitemShut
  {NoStop}%
\bibitem [{\citenamefont {Bitzek}\ \emph {et~al.}(2006)\citenamefont {Bitzek},
  \citenamefont {Koskinen}, \citenamefont {G\"ahler}, \citenamefont {Moseler},\
  and\ \citenamefont {Gumbsch}}]{fire}%
  \BibitemOpen
  \bibfield  {author} {\bibinfo {author} {\bibfnamefont {E.}~\bibnamefont
  {Bitzek}}, \bibinfo {author} {\bibfnamefont {P.}~\bibnamefont {Koskinen}},
  \bibinfo {author} {\bibfnamefont {F.}~\bibnamefont {G\"ahler}}, \bibinfo
  {author} {\bibfnamefont {M.}~\bibnamefont {Moseler}}, \ and\ \bibinfo
  {author} {\bibfnamefont {P.}~\bibnamefont {Gumbsch}},\ }\href {\doibase
  10.1103/PhysRevLett.97.170201} {\bibfield  {journal} {\bibinfo  {journal}
  {Phys. Rev. Lett.}\ }\textbf {\bibinfo {volume} {97}},\ \bibinfo {pages}
  {170201} (\bibinfo {year} {2006})}\BibitemShut {NoStop}%
\bibitem [{\citenamefont {Berendsen}\ \emph {et~al.}(1984)\citenamefont
  {Berendsen}, \citenamefont {Postma}, \citenamefont {van Gunsteren},
  \citenamefont {DiNola},\ and\ \citenamefont {Haak}}]{berendsen}%
  \BibitemOpen
  \bibfield  {author} {\bibinfo {author} {\bibfnamefont {H.~J.~C.}\
  \bibnamefont {Berendsen}}, \bibinfo {author} {\bibfnamefont {J.~P.~M.}\
  \bibnamefont {Postma}}, \bibinfo {author} {\bibfnamefont {W.~F.}\
  \bibnamefont {van Gunsteren}}, \bibinfo {author} {\bibfnamefont
  {A.}~\bibnamefont {DiNola}}, \ and\ \bibinfo {author} {\bibfnamefont {J.~R.}\
  \bibnamefont {Haak}},\ }\href {\doibase http://dx.doi.org/10.1063/1.448118}
  {\bibfield  {journal} {\bibinfo  {journal} {J. Chem. Phys.}\ }\textbf
  {\bibinfo {volume} {81}},\ \bibinfo {pages} {3684} (\bibinfo {year}
  {1984})}\BibitemShut {NoStop}%
\bibitem [{\citenamefont {O'Hern}\ \emph {et~al.}(2003)\citenamefont {O'Hern},
  \citenamefont {Silbert}, \citenamefont {Liu},\ and\ \citenamefont
  {Nagel}}]{ohern2003}%
  \BibitemOpen
  \bibfield  {author} {\bibinfo {author} {\bibfnamefont {C.~S.}\ \bibnamefont
  {O'Hern}}, \bibinfo {author} {\bibfnamefont {L.~E.}\ \bibnamefont {Silbert}},
  \bibinfo {author} {\bibfnamefont {A.~J.}\ \bibnamefont {Liu}}, \ and\
  \bibinfo {author} {\bibfnamefont {S.~R.}\ \bibnamefont {Nagel}},\ }\href
  {\doibase 10.1103/PhysRevE.68.011306} {\bibfield  {journal} {\bibinfo
  {journal} {Phys. Rev. E}\ }\textbf {\bibinfo {volume} {68}},\ \bibinfo
  {pages} {011306} (\bibinfo {year} {2003})}\BibitemShut {NoStop}%
\bibitem [{\citenamefont {DeGiuli}\ \emph {et~al.}(2014)\citenamefont
  {DeGiuli}, \citenamefont {Laversanne-Finot}, \citenamefont {During},
  \citenamefont {Lerner},\ and\ \citenamefont {Wyart}}]{eric_boson_peak_emt}%
  \BibitemOpen
  \bibfield  {author} {\bibinfo {author} {\bibfnamefont {E.}~\bibnamefont
  {DeGiuli}}, \bibinfo {author} {\bibfnamefont {A.}~\bibnamefont
  {Laversanne-Finot}}, \bibinfo {author} {\bibfnamefont {G.}~\bibnamefont
  {During}}, \bibinfo {author} {\bibfnamefont {E.}~\bibnamefont {Lerner}}, \
  and\ \bibinfo {author} {\bibfnamefont {M.}~\bibnamefont {Wyart}},\ }\href
  {\doibase 10.1039/C4SM00561A} {\bibfield  {journal} {\bibinfo  {journal}
  {Soft Matter}\ }\textbf {\bibinfo {volume} {10}},\ \bibinfo {pages} {5628}
  (\bibinfo {year} {2014})}\BibitemShut {NoStop}%
\bibitem [{\citenamefont {Franz}\ \emph {et~al.}(2015)\citenamefont {Franz},
  \citenamefont {Parisi}, \citenamefont {Urbani},\ and\ \citenamefont
  {Zamponi}}]{silvio}%
  \BibitemOpen
  \bibfield  {author} {\bibinfo {author} {\bibfnamefont {S.}~\bibnamefont
  {Franz}}, \bibinfo {author} {\bibfnamefont {G.}~\bibnamefont {Parisi}},
  \bibinfo {author} {\bibfnamefont {P.}~\bibnamefont {Urbani}}, \ and\ \bibinfo
  {author} {\bibfnamefont {F.}~\bibnamefont {Zamponi}},\ }\href {\doibase
  10.1073/pnas.1511134112} {\bibfield  {journal} {\bibinfo  {journal} {Proc.
  Natl. Acad. Sci. U.S.A.}\ }\textbf {\bibinfo {volume} {112}},\ \bibinfo
  {pages} {14539} (\bibinfo {year} {2015})}\BibitemShut {NoStop}%
\bibitem [{\citenamefont {Mizuno}\ \emph {et~al.}(2017)\citenamefont {Mizuno},
  \citenamefont {Shiba},\ and\ \citenamefont {Ikeda}}]{ikeda_pnas}%
  \BibitemOpen
  \bibfield  {author} {\bibinfo {author} {\bibfnamefont {H.}~\bibnamefont
  {Mizuno}}, \bibinfo {author} {\bibfnamefont {H.}~\bibnamefont {Shiba}}, \
  and\ \bibinfo {author} {\bibfnamefont {A.}~\bibnamefont {Ikeda}},\ }\href
  {\doibase 10.1073/pnas.1709015114} {\bibfield  {journal} {\bibinfo  {journal}
  {Proc. Natl. Acad. Sci. U.S.A.}\ }\textbf {\bibinfo {volume} {114}},\
  \bibinfo {pages} {E9767} (\bibinfo {year} {2017})}\BibitemShut {NoStop}%
\bibitem [{\citenamefont {Gartner}\ and\ \citenamefont
  {Lerner}(2016)}]{plastic_modes_prerc}%
  \BibitemOpen
  \bibfield  {author} {\bibinfo {author} {\bibfnamefont {L.}~\bibnamefont
  {Gartner}}\ and\ \bibinfo {author} {\bibfnamefont {E.}~\bibnamefont
  {Lerner}},\ }\href {\doibase 10.1103/PhysRevE.93.011001} {\bibfield
  {journal} {\bibinfo  {journal} {Phys. Rev. E}\ }\textbf {\bibinfo {volume}
  {93}},\ \bibinfo {pages} {011001} (\bibinfo {year} {2016})}\BibitemShut
  {NoStop}%
\bibitem [{\citenamefont {Ji}\ \emph {et~al.}(2018)\citenamefont {Ji},
  \citenamefont {Popovi\'c}, \citenamefont {W.~J.~de Geus}, \citenamefont
  {Lerner},\ and\ \citenamefont {Wyart}}]{wencheng}%
  \BibitemOpen
  \bibfield  {author} {\bibinfo {author} {\bibfnamefont {W.}~\bibnamefont
  {Ji}}, \bibinfo {author} {\bibfnamefont {M.}~\bibnamefont {Popovi\'c}},
  \bibinfo {author} {\bibfnamefont {T.}~\bibnamefont {W.~J.~de Geus}}, \bibinfo
  {author} {\bibfnamefont {E.}~\bibnamefont {Lerner}}, \ and\ \bibinfo {author}
  {\bibfnamefont {M.}~\bibnamefont {Wyart}},\ }\href
  {https://arxiv.org/abs/1806.01561} {\bibfield  {journal} {\bibinfo  {journal}
  {arXiv preprint arXiv:1806.01561}\ } (\bibinfo {year} {2018})}\BibitemShut
  {NoStop}%
\bibitem [{\citenamefont {Hentschel}\ \emph {et~al.}(2011)\citenamefont
  {Hentschel}, \citenamefont {Karmakar}, \citenamefont {Lerner},\ and\
  \citenamefont {Procaccia}}]{exist}%
  \BibitemOpen
  \bibfield  {author} {\bibinfo {author} {\bibfnamefont {H.~G.~E.}\
  \bibnamefont {Hentschel}}, \bibinfo {author} {\bibfnamefont {S.}~\bibnamefont
  {Karmakar}}, \bibinfo {author} {\bibfnamefont {E.}~\bibnamefont {Lerner}}, \
  and\ \bibinfo {author} {\bibfnamefont {I.}~\bibnamefont {Procaccia}},\ }\href
  {\doibase 10.1103/PhysRevE.83.061101} {\bibfield  {journal} {\bibinfo
  {journal} {Phys. Rev. E}\ }\textbf {\bibinfo {volume} {83}},\ \bibinfo
  {pages} {061101} (\bibinfo {year} {2011})}\BibitemShut {NoStop}%
\bibitem [{\citenamefont {Karpov}\ and\ \citenamefont
  {Parshin}(1985)}]{soft_potential_model_02}%
  \BibitemOpen
  \bibfield  {author} {\bibinfo {author} {\bibfnamefont {V.}~\bibnamefont
  {Karpov}}\ and\ \bibinfo {author} {\bibfnamefont {D.}~\bibnamefont
  {Parshin}},\ }\href@noop {} {\bibfield  {journal} {\bibinfo  {journal} {Zh.
  Eksp. Teor. Fiz}\ }\textbf {\bibinfo {volume} {88}},\ \bibinfo {pages} {2212}
  (\bibinfo {year} {1985})}\BibitemShut {NoStop}%
\bibitem [{\citenamefont {Buchenau}\ \emph {et~al.}(1991)\citenamefont
  {Buchenau}, \citenamefont {Galperin}, \citenamefont {Gurevich},\ and\
  \citenamefont {Schober}}]{soft_potential_model_1991}%
  \BibitemOpen
  \bibfield  {author} {\bibinfo {author} {\bibfnamefont {U.}~\bibnamefont
  {Buchenau}}, \bibinfo {author} {\bibfnamefont {Y.~M.}\ \bibnamefont
  {Galperin}}, \bibinfo {author} {\bibfnamefont {V.~L.}\ \bibnamefont
  {Gurevich}}, \ and\ \bibinfo {author} {\bibfnamefont {H.~R.}\ \bibnamefont
  {Schober}},\ }\href {\doibase 10.1103/PhysRevB.43.5039} {\bibfield  {journal}
  {\bibinfo  {journal} {Phys. Rev. B}\ }\textbf {\bibinfo {volume} {43}},\
  \bibinfo {pages} {5039} (\bibinfo {year} {1991})}\BibitemShut {NoStop}%
\bibitem [{\citenamefont {Buchenau}\ \emph {et~al.}(1992)\citenamefont
  {Buchenau}, \citenamefont {Galperin}, \citenamefont {Gurevich}, \citenamefont
  {Parshin}, \citenamefont {Ramos},\ and\ \citenamefont
  {Schober}}]{Buchenau_1992}%
  \BibitemOpen
  \bibfield  {author} {\bibinfo {author} {\bibfnamefont {U.}~\bibnamefont
  {Buchenau}}, \bibinfo {author} {\bibfnamefont {Y.~M.}\ \bibnamefont
  {Galperin}}, \bibinfo {author} {\bibfnamefont {V.~L.}\ \bibnamefont
  {Gurevich}}, \bibinfo {author} {\bibfnamefont {D.~A.}\ \bibnamefont
  {Parshin}}, \bibinfo {author} {\bibfnamefont {M.~A.}\ \bibnamefont {Ramos}},
  \ and\ \bibinfo {author} {\bibfnamefont {H.~R.}\ \bibnamefont {Schober}},\
  }\href {\doibase 10.1103/PhysRevB.46.2798} {\bibfield  {journal} {\bibinfo
  {journal} {Phys. Rev. B}\ }\textbf {\bibinfo {volume} {46}},\ \bibinfo
  {pages} {2798} (\bibinfo {year} {1992})}\BibitemShut {NoStop}%
\bibitem [{\citenamefont {Baity-Jesi}\ \emph {et~al.}(2015)\citenamefont
  {Baity-Jesi}, \citenamefont {Mart\'{\i}n-Mayor}, \citenamefont {Parisi},\
  and\ \citenamefont {Perez-Gaviro}}]{parisi_spin_glass}%
  \BibitemOpen
  \bibfield  {author} {\bibinfo {author} {\bibfnamefont {M.}~\bibnamefont
  {Baity-Jesi}}, \bibinfo {author} {\bibfnamefont {V.}~\bibnamefont
  {Mart\'{\i}n-Mayor}}, \bibinfo {author} {\bibfnamefont {G.}~\bibnamefont
  {Parisi}}, \ and\ \bibinfo {author} {\bibfnamefont {S.}~\bibnamefont
  {Perez-Gaviro}},\ }\href {\doibase 10.1103/PhysRevLett.115.267205} {\bibfield
   {journal} {\bibinfo  {journal} {Phys. Rev. Lett.}\ }\textbf {\bibinfo
  {volume} {115}},\ \bibinfo {pages} {267205} (\bibinfo {year}
  {2015})}\BibitemShut {NoStop}%
\bibitem [{\citenamefont {Lutsko}(1989)}]{lutsko}%
  \BibitemOpen
  \bibfield  {author} {\bibinfo {author} {\bibfnamefont {J.~F.}\ \bibnamefont
  {Lutsko}},\ }\href {\doibase http://dx.doi.org/10.1063/1.342716} {\bibfield
  {journal} {\bibinfo  {journal} {J. Appl. Phys.}\ }\textbf {\bibinfo {volume}
  {65}},\ \bibinfo {pages} {2991} (\bibinfo {year} {1989})}\BibitemShut
  {NoStop}%
\bibitem [{Note1()}]{Note1}%
  \BibitemOpen
  \bibinfo {note} {We thank Misaki Ozawa for pointing out this
  detail.}\BibitemShut {Stop}%
\bibitem [{\citenamefont {Jin}\ \emph {et~al.}(2018)\citenamefont {Jin},
  \citenamefont {Urbani}, \citenamefont {Zamponi},\ and\ \citenamefont
  {Yoshino}}]{Yoshino2018}%
  \BibitemOpen
  \bibfield  {author} {\bibinfo {author} {\bibfnamefont {Y.}~\bibnamefont
  {Jin}}, \bibinfo {author} {\bibfnamefont {P.}~\bibnamefont {Urbani}},
  \bibinfo {author} {\bibfnamefont {F.}~\bibnamefont {Zamponi}}, \ and\
  \bibinfo {author} {\bibfnamefont {H.}~\bibnamefont {Yoshino}},\ }\href
  {\doibase 10.1126/sciadv.aat6387} {\bibfield  {journal} {\bibinfo  {journal}
  {Science Advances}\ }\textbf {\bibinfo {volume} {4}} (\bibinfo {year}
  {2018}),\ 10.1126/sciadv.aat6387}\BibitemShut {NoStop}%
\bibitem [{Note2()}]{Note2}%
  \BibitemOpen
  \bibinfo {note} {We consider the lowest vibrational mode per glass in systems
  of $N\protect \tmspace -\thinmuskip {.1667em}=\protect \tmspace -\thinmuskip
  {.1667em}4000$ particles, which is typically quasilocalized \cite
  {modes_prl}}\BibitemShut {NoStop}%
\bibitem [{\citenamefont {Shimada}\ \emph {et~al.}(2018)\citenamefont
  {Shimada}, \citenamefont {Mizuno}, \citenamefont {Wyart},\ and\ \citenamefont
  {Ikeda}}]{Ikeda2018}%
  \BibitemOpen
  \bibfield  {author} {\bibinfo {author} {\bibfnamefont {M.}~\bibnamefont
  {Shimada}}, \bibinfo {author} {\bibfnamefont {H.}~\bibnamefont {Mizuno}},
  \bibinfo {author} {\bibfnamefont {M.}~\bibnamefont {Wyart}}, \ and\ \bibinfo
  {author} {\bibfnamefont {A.}~\bibnamefont {Ikeda}},\ }\href
  {https://arxiv.org/abs/1804.08865} {\bibfield  {journal} {\bibinfo  {journal}
  {arXiv preprint arXiv:1804.08865}\ } (\bibinfo {year} {2018})}\BibitemShut
  {NoStop}%
\bibitem [{\citenamefont {Mizuno}\ \emph {et~al.}(2016)\citenamefont {Mizuno},
  \citenamefont {Saitoh},\ and\ \citenamefont {Silbert}}]{Silbert2016PRE}%
  \BibitemOpen
  \bibfield  {author} {\bibinfo {author} {\bibfnamefont {H.}~\bibnamefont
  {Mizuno}}, \bibinfo {author} {\bibfnamefont {K.}~\bibnamefont {Saitoh}}, \
  and\ \bibinfo {author} {\bibfnamefont {L.~E.}\ \bibnamefont {Silbert}},\
  }\href {\doibase 10.1103/PhysRevE.93.062905} {\bibfield  {journal} {\bibinfo
  {journal} {Phys. Rev. E}\ }\textbf {\bibinfo {volume} {93}},\ \bibinfo
  {pages} {062905} (\bibinfo {year} {2016})}\BibitemShut {NoStop}%
\bibitem [{\citenamefont {Alexander}(1998)}]{shlomo}%
  \BibitemOpen
  \bibfield  {author} {\bibinfo {author} {\bibfnamefont {S.}~\bibnamefont
  {Alexander}},\ }\href {\doibase
  http://dx.doi.org/10.1016/S0370-1573(97)00069-0} {\bibfield  {journal}
  {\bibinfo  {journal} {Phys. Rep.}\ }\textbf {\bibinfo {volume} {296}},\
  \bibinfo {pages} {65 } (\bibinfo {year} {1998})}\BibitemShut {NoStop}%
\bibitem [{\citenamefont {Wyart}\ \emph {et~al.}(2005)\citenamefont {Wyart},
  \citenamefont {Silbert}, \citenamefont {Nagel},\ and\ \citenamefont
  {Witten}}]{matthieu_PRE_2005}%
  \BibitemOpen
  \bibfield  {author} {\bibinfo {author} {\bibfnamefont {M.}~\bibnamefont
  {Wyart}}, \bibinfo {author} {\bibfnamefont {L.~E.}\ \bibnamefont {Silbert}},
  \bibinfo {author} {\bibfnamefont {S.~R.}\ \bibnamefont {Nagel}}, \ and\
  \bibinfo {author} {\bibfnamefont {T.~A.}\ \bibnamefont {Witten}},\ }\href
  {\doibase 10.1103/PhysRevE.72.051306} {\bibfield  {journal} {\bibinfo
  {journal} {Phys. Rev. E}\ }\textbf {\bibinfo {volume} {72}},\ \bibinfo
  {pages} {051306} (\bibinfo {year} {2005})}\BibitemShut {NoStop}%
\bibitem [{\citenamefont {Wyart}(2005)}]{matthieu_thesis}%
  \BibitemOpen
  \bibfield  {author} {\bibinfo {author} {\bibfnamefont {M.}~\bibnamefont
  {Wyart}},\ }in\ \href {\doibase https://doi.org/10.1051/anphys:2006003}
  {\emph {\bibinfo {booktitle} {Annales de Physique}}},\ Vol.~\bibinfo {volume}
  {30}\ (\bibinfo {year} {2005})\ pp.\ \bibinfo {pages} {1--96}\BibitemShut
  {NoStop}%
\bibitem [{\citenamefont {Xu}\ \emph {et~al.}(2007)\citenamefont {Xu},
  \citenamefont {Wyart}, \citenamefont {Liu},\ and\ \citenamefont
  {Nagel}}]{Xu07}%
  \BibitemOpen
  \bibfield  {author} {\bibinfo {author} {\bibfnamefont {N.}~\bibnamefont
  {Xu}}, \bibinfo {author} {\bibfnamefont {M.}~\bibnamefont {Wyart}}, \bibinfo
  {author} {\bibfnamefont {A.~J.}\ \bibnamefont {Liu}}, \ and\ \bibinfo
  {author} {\bibfnamefont {S.~R.}\ \bibnamefont {Nagel}},\ }\href {\doibase
  10.1103/PhysRevLett.98.175502} {\bibfield  {journal} {\bibinfo  {journal}
  {Phys. Rev. Lett.}\ }\textbf {\bibinfo {volume} {98}},\ \bibinfo {pages}
  {175502} (\bibinfo {year} {2007})}\BibitemShut {NoStop}%
\bibitem [{\citenamefont {P\'erez-Reche}\ \emph {et~al.}(2008)\citenamefont
  {P\'erez-Reche}, \citenamefont {Truskinovsky},\ and\ \citenamefont
  {Zanzotto}}]{Perez08}%
  \BibitemOpen
  \bibfield  {author} {\bibinfo {author} {\bibfnamefont {F.-J.}\ \bibnamefont
  {P\'erez-Reche}}, \bibinfo {author} {\bibfnamefont {L.}~\bibnamefont
  {Truskinovsky}}, \ and\ \bibinfo {author} {\bibfnamefont {G.}~\bibnamefont
  {Zanzotto}},\ }\href {\doibase 10.1103/PhysRevLett.101.230601} {\bibfield
  {journal} {\bibinfo  {journal} {Phys. Rev. Lett.}\ }\textbf {\bibinfo
  {volume} {101}},\ \bibinfo {pages} {230601} (\bibinfo {year}
  {2008})}\BibitemShut {NoStop}%
\bibitem [{\citenamefont {Popovi\ifmmode~\acute{c}\else \'{c}\fi{}}\ \emph
  {et~al.}(2018)\citenamefont {Popovi\ifmmode~\acute{c}\else \'{c}\fi{}},
  \citenamefont {de~Geus},\ and\ \citenamefont {Wyart}}]{Popovic2018}%
  \BibitemOpen
  \bibfield  {author} {\bibinfo {author} {\bibfnamefont {M.}~\bibnamefont
  {Popovi\ifmmode~\acute{c}\else \'{c}\fi{}}}, \bibinfo {author} {\bibfnamefont
  {T.~W.~J.}\ \bibnamefont {de~Geus}}, \ and\ \bibinfo {author} {\bibfnamefont
  {M.}~\bibnamefont {Wyart}},\ }\href {\doibase 10.1103/PhysRevE.98.040901}
  {\bibfield  {journal} {\bibinfo  {journal} {Phys. Rev. E}\ }\textbf {\bibinfo
  {volume} {98}},\ \bibinfo {pages} {040901} (\bibinfo {year}
  {2018})}\BibitemShut {NoStop}%
\bibitem [{\citenamefont {Lehoucq}\ \emph {et~al.}(1998)\citenamefont
  {Lehoucq}, \citenamefont {Sorensen},\ and\ \citenamefont {Yang}}]{arpack}%
  \BibitemOpen
  \bibfield  {author} {\bibinfo {author} {\bibfnamefont {R.~B.}\ \bibnamefont
  {Lehoucq}}, \bibinfo {author} {\bibfnamefont {D.~C.}\ \bibnamefont
  {Sorensen}}, \ and\ \bibinfo {author} {\bibfnamefont {C.}~\bibnamefont
  {Yang}},\ }\href {\doibase 10.1137/1.9780898719628} {\emph {\bibinfo {title}
  {ARPACK users' guide: solution of large-scale eigenvalue problems with
  implicitly restarted Arnoldi methods}}},\ Vol.~\bibinfo {volume} {6}\
  (\bibinfo  {publisher} {Siam},\ \bibinfo {year} {1998})\BibitemShut {NoStop}%
\bibitem [{\citenamefont {Allen}\ and\ \citenamefont
  {Tildesley}(1989)}]{allen1989computer}%
  \BibitemOpen
  \bibfield  {author} {\bibinfo {author} {\bibfnamefont {M.~P.}\ \bibnamefont
  {Allen}}\ and\ \bibinfo {author} {\bibfnamefont {D.~J.}\ \bibnamefont
  {Tildesley}},\ }\href@noop {} {\emph {\bibinfo {title} {Computer simulation
  of liquids}}}\ (\bibinfo  {publisher} {Oxford university press},\ \bibinfo
  {year} {1989})\BibitemShut {NoStop}%
\end{thebibliography}%

\end{document}